\documentclass[iop, twocolappendix, appendixfloats, numberedappendix, apj]{emulateapj}

\usepackage{graphicx}
\usepackage{verbatim}
\usepackage{color}
\usepackage{xcolor}
\usepackage{amsmath} 
\usepackage{enumitem}
\usepackage[colorlinks=true, citecolor=blue, urlcolor=blue, linkcolor=blue]{hyperref}

\def\eps@scaling{1.0}%

\newcommand{\degrade}{20\%}

\newcommand{\snr}{$S/N$}
\newcommand{\hlr}{$r_{50}$}

\newcommand{\vecg}{\mbox{\boldmath $\gamma$}}
\newcommand{\vest}{\mbox{\boldmath $e$}}
\newcommand{\est}{e}

\newcommand{\mcal}{\textsc{metacalibration}}
\newcommand{\Mcal}{\textsc{Metacalibration}}

\newcommand{\psfex}{\texttt{PSFEx}}

\newcommand{\mcalR}{\mbox{\boldmath $R$}}

\newcommand{\mcalRmean}{\mbox{\boldmath $\langle R \rangle$}}

\newcommand{\mcalRo}{\mbox{\boldmath $R_o$}}
\newcommand{\mcalRnoise}{\mbox{\boldmath $R_\eta$}}

\newcommand{\mcalRmeanalpha}{\mbox{\boldmath $\langle R_\alpha \rangle$}}
\newcommand{\mcalRmeanbeta}{\mbox{\boldmath $\langle R_\beta \rangle$}}

\newcommand{\mcalRg}{\mbox{\boldmath $R_\gamma$}}
\newcommand{\mcalRS}{\mbox{\boldmath $R_S$}}
\newcommand{\mcalRgmean}{\mbox{\boldmath $\langle R_\gamma \rangle$}}
\newcommand{\mcalRSmean}{\mbox{\boldmath $\langle R_S \rangle$}}

\newcommand{\mcalRtwopt}{\mbox{\boldmath $R^{2pt}$}}
\newcommand{\mcalRtwoptmean}{\mbox{\boldmath $\langle R^{2pt} \rangle$}}

\newcommand{\mcalRmodel}{\mbox{\boldmath $R^{model}$}}
\newcommand{\mcalRnoisemodel}{\mbox{\boldmath $R^{model}_\eta$}}

\newcommand{\probe}{\mbox{$P(\vest_\alpha, \vest_\beta)$}}

\newcommand{\Itild}{\mbox{$\widetilde{I}$}}
\newcommand{\ntil}{\mbox{$\widetilde{\eta}$}}
\newcommand{\Ptil}{\mbox{$\widetilde{P}$}}
\newcommand{\Ptild}{\mbox{$\widetilde{P_d}$}}

\newcommand{\nsimShear}{0.02,0.00}
\newcommand{\nsimNgal}{$10^8$}

\newcommand{\nsimNgalBD}{$5.6 \times 10^9$}
\newcommand{\nsimNStar}{$5.6 \times 10^8$}
\newcommand{\nsimNstarperc}{10\%}
\newcommand{\starnoiseincrease}{$\sim 2-3$\%}

\newcommand{\nsimPSFShape}{$0.000,0.025$}

\newcommand{\psfrdist}{0.9}

\newcommand{\galrdist}{COSMOS Fits}

\newcommand{\cosmosname}{COSMOS}

\newcommand{\detrend}{detrending}
\newcommand{\Detrend}{Detrending}

\newcommand{\fixnoise}{\texttt{fixnoise}}

\newcommand{\ngmix}{\texttt{ngmix}}

\newcommand{\Aslope}{$-1.11$}
\newcommand{\Rcorr}{$-0.0694$}

\newcommand{\bdkfull}{Bulge+Disk+Knots}
\newcommand{\bdksim}{\texttt{BDK}}
\newcommand{\bdstar}{\texttt{BDK+Stars}}

\newcommand{\rgsim}{\texttt{RG}}

\newcommand{\galsim}{\texttt{GALSIM}}
\newcommand{\balrog}{\texttt{BALROG}}
\newcommand{\bfd}{\texttt{BFD}}

\newcommand{\RR}{$\left<RR\right>$}
\newcommand{\RS}{$\left<RR_{S}\right>$}
\newcommand{\SSs}{$\left<R_{S}R_{S}\right>$}

\shorttitle{Practical Weak Lensing Shear Measurement with Metacalibration}
\shortauthors{Sheldon and Huff}

\begin{document}

\title{Practical Weak Lensing Shear Measurement with Metacalibration}

\author{Erin S. Sheldon}
\affil{Brookhaven National Laboratory, Bldg 510, Upton, New York 11973}
\and
\author{Eric M. Huff}
\affil{Jet  Propulsion  Laboratory,  California  Institute  of  Technology, 4800 Oak Grove Dr., Pasadena, CA 91109, USA}

 

\keywords{
    cosmology: observations --- gravitational lensing: weak --- methods: observational
}

\begin{abstract}

    \Mcal\ is a recently introduced method to accurately measure weak
    gravitational lensing shear using only the available imaging data, without
    need for prior information about galaxy properties or calibration from
    simulations.  The method involves distorting the image with a small known
    shear, and calculating the response of a shear estimator to that applied
    shear.  The method was shown to be accurate in moderate sized simulations
    with galaxy images that had relatively high signal-to-noise ratios, and
    without significant selection effects.  In this work we introduce a
    formalism to correct for both shear response and selection biases.  We also
    observe that, for images with relatively low signal-to-noise ratios,  the
    correlated noise that arises during the \mcal\ process results in
    significant bias, for which we develop a simple empirical correction.  To
    test this formalism, we created large image simulations based on both
    parametric models and real galaxy images, including tests with realistic
    point-spread functions.  We varied the point-spread function ellipticity at
    the five percent level.  In each simulation we applied a small, few percent
    shear to the galaxy images.  We introduced additional challenges that arise
    in real data, such as detection thresholds, stellar contamination, and
    missing data.  We applied cuts on the measured galaxy properties to induce
    significant selection effects.  Using our formalism, we recovered the input
    shear with an accuracy better than a part in a thousand in all cases.

\end{abstract}


\section{Introduction} \label{sec:intro}

Weak gravitational lensing is a fascinating phenomenon that has become a useful
tool for testing our theories of gravity, measuring the properties and
distribution of dark matter, and characterizing the accelerated expansion 
of the universe known as dark energy \citep[for a review, see][]{HoekstraJain2008}.

Light passing massive objects undergoes a deflection, and the amount of this
deflection depends on all the mass in the ``lens'', both luminous and dark.
The deflection distorts, or ``shears'', the observed shape of extended objects
such as galaxies, and this shear is spatially correlated.  By studying these
spatial correlations, one can infer the correlations in the mass that caused the
lensing. One can study these correlations at any place in the universe,
including the locations of galaxies \citep{Mandelbaum06}, clusters of galaxies
\citep{JohnstonLensing07}, voids \citep{MelchiorVoids2014}, and even
correlations without reference to any particular lensing objects
\citep{CFHTCosmicShear2013}.  Because the effect depends on the geometry of the
lens-source system as well as mass, the expansion history and growth of
structure can be inferred, and thus weak lensing is also sensitive to dark
energy \citep{HeymansTomography2013}.

Pioneered as a measurement technique in the 1980s and 1990s, weak lensing was
quickly recognized to be complimentary to more standard dynamical techniques
\citep{Tyson84}.  Because the effect is independent of the dynamical state of
the lensing mass, it can be used to study the distribution of mass in systems
that are far from equilibrium \citep{CloweDMProof06}.   And because the effect
can be measured nearly anywhere, it can be used to probe very large scales,
measuring the correlation between objects and large scale structure
\citep{SheldonLensing09}.

Thus far, progress in measurement has been limited by technical challenges, a
few of which we will outline below.  To set a scale of reference, the weak
lensing shear due to a foreground lens typically introduces correlations in the
shapes of background galaxies at the percent level.    We would like to measure
this percent level signal to approximately a part in a thousand in upcoming
imaging surveys \citep{HutererSystematics06}.

There have traditionally been two approaches to measuring shear from
galaxy images: measuring moments of the light distribution and fitting models.

In the model fitting approach, a parametric model is convolved by an estimate
of the point-spread function (PSF) of the atmosphere and instrument, and fit to
the galaxy surface brightness profile.  The shear estimator, the ``shape'' of
the galaxy, is then determined from the parameters of this model.  This method
is limited for two primary reasons: first, in order for the shear inference to
be accurate, the model must accurately reproduce the galaxy profile before
noise and convolution by the PSF, which requires a large number of parameters.
If the model cannot sufficiently reproduce the galaxy the estimator is said to
exhibit ``model bias'' \citep{Bernstein2010}.  The second is the bias in the
nonlinear model fitting caused by noise, which is significant for images with low
signal-to-noise ratios (\snr), and is difficult to predict
\citep{HirataAlign04,Refreg12,Melchior12}.  This ``noise bias'' is made worse
if large number of model parameters are used in an attempt to accurately
represent the galaxy.

Because of these biases, model fitting methods require additional
calibration.  However, there are no absolute calibration sources in the
universe that can be used to derive a calibration.  Instead, image
simulations have often been used to determine the shear calibration
\citep[e.g.][]{Zuntz13,Miller13,KidsShear2017,Refregier13,Jee16}.  These
simulations must include all the relevant details of the real universe in order
to provide an accurate calibration.  It has been suggested that model biases
can partly be alleviated without image simulations if a sufficiently flexible
statistical framework is used \citep{SchneiderProbshear2015}, but this has not
yet been demonstrated.

Another traditional shear measurement technique uses the second moments of the
galaxy light distribution, after correction for the effect of the PSF
\citep[e.g.][]{ksb95,Bernstein2010}.  These methods can be made quite accurate
for galaxy images with high \snr, with little model bias
\citep{Bernstein2010,Okura2016}.  However, unlike model fitting approaches,
missing data and light from nearby objects cannot be ``masked-out'' or ignored.
Also, these methods still suffer the ``noise bias'' found in model fitting
methods as a result of the nonlinear fitting involved with the measurement.
Without further development, these methods are not accurate at the
part-in-a-thousand level.

Finally, selection effects can bias the recovered shear in both model fitting
and moment based methods, but the corrections are typically considered as
separate from the shear estimation, to be inferred as part of a calibration
based on simulations \citep{Jarvis2016,KidsShear2017}).

Methods have recently been developed to avoid some of these biases without
relying on calibration from simulations.  The method of \cite{Zhang2017}
involves measuring the weighted moments in Fourier space, after subtracting an
appropriate noise power spectrum to deal with noise effects, and deconvolving
the PSF.  No nonlinear fitting is performed.  The shear is then estimated
either from the ratio of sums of these moments \citep{Zhang2015}, or the PDF of
the un-normalized moments is symmetrized in order to infer an applied shear
\citep{Zhang2017}.  The sum method was shown to be accurate at the
part-in-a-thousand level in challenging simulations, but with unacceptably high
noise.  The PDF symmetrization method was shown to be more precise, but the
simulations used were not large enough to determine the accuracy of the method
to the part-in-a-thousand level.  No formalism has been proposed as yet to deal
with selection effects. 

Another new approach is the Bayesian Fourier Domain (\bfd) method for shear
inference introduced by \cite{ba14} and further developed by \cite{bfd2016}.  A
rigorous Bayesian framework was presented, in which prior information on galaxy
images from much deeper data is included.  Un-normalized moments in
Fourier space are used as the basic data vector, avoiding nonlinear fitting
issues.  \bfd\ is also the first method for which selection effects are
corrected for naturally in the formalism, without use of external
simulations.  The method was tested in \cite{bfd2016} using a challenging
simulation.  In that initial study, a bias of $\sim 2 \times 10^{-3}$ was
detected, falling just short of the part-in-a-thousand target.  However, given
its rigorous foundation, it seems possible that the desired accuracy will be
achieved with further technical development.  The required prior information
from deep data should in principle be available to current and future surveys
\citep{DESWhitePaper,TakadaHSC2010,IvezicLSST08,Euclid2011,SpergelWFIRST2015}.


\Mcal\ \citep{HuffMcal}, the subject of this work, is a new method designed to
calibrate standard shear estimators, without requiring significant prior
information about galaxy properties.  This is accomplished by introducing an
artificial shear to images, and calculating how the shear estimator responds to
the applied shear, addressing both model bias and noise bias.  \Mcal\ can in
principle be used to calibrate any shear estimator, including shapes derived
from model fitting or weighted moments.

The \mcal\ approach is not entirely new. A quite similar idea to \mcal\ was
introduced by \cite{Kaiser2000}, although the full equivalent of \mcal\ was not
implemented therein.  \cite{ksb95} also introduced the use of sheared
high-resolution space-based images as a way to provide an overall calibration
for a shear estimator.


The \mcal\ technique was shown to be accurate in controlled simulations
\citep{HuffMcal}.  However, the simulations used in that study \citep[based on
those used in][]{great3} contained galaxy images with fairly high \snr, and the
galaxies were relatively large compared to the PSF size.  In a real survey, the
images are dominated by small faint galaxies, many of which cannot be reliably
detected or measured.  Furthermore, owing to the limited number of galaxies in
those simulations, the accuracy of the method could only be tested to about
three parts in a thousand.  In this work we use a more challenging set of
simulations.

We will also address other difficulties that arise in real data.  During
estimation of a shear statistic, further selection beyond a simple detection
threshold is often required, such as cutting or binning based on galaxy
properties, which can cause significant selection biases.  We introduce a
formalism to deal with selection effects, for example, cuts on \snr\ and galaxy
size, that are required for any practical analysis. 

In real images, stars cannot be perfectly removed from the sample of objects
used to measure shear. We will show that \mcal\ is robust to the presence of
stars in the sample if the PSF is well determined.


Finally, the \mcal\ procedure itself correlates the noise across the image,
which becomes a dominant source of error for galaxy images with low \snr.  We
introduce a simple image-level correction for this correlated noise.

We show that, in all scenarios we tested, \mcal\ is indeed accurate to a part
in a thousand.

\section{Introduction to Metacalibration} \label{sec:mcal}

In this section we introduce the basic concepts of \mcal. We derive the full
formalism for shear measurements in \S \ref{sec:formalism}.

Suppose we have a noisy measurement \vest\ that we wish to use for shear
estimation.  \vest\ may be some estimate of an objects two-component ellipticity
such that $\vest = (\est_1, \est_2)$.  We can expand this observable in a
Taylor series about zero shear
\begin{align} \label{eq:Eexpand}
    \vest &= \vest|_{\gamma=0} + \frac{ \partial \vest }{ \partial \vecg}\bigg|_{\gamma=0} \vecg  + ... \nonumber \\
          &\equiv \vest|_{\gamma=0} + \mbox{\mcalR}\vecg  + ...
\end{align}
where we have defined the shear response:
\begin{align}
    \mbox{\mcalR} &\equiv \frac{\partial \vest}{\partial \vecg} \bigg|_{\gamma=0}.
\end{align}
Note that the derivative is also with respect to the two-component shear $\vecg$, making
\mcalR\ a $2 \times 2$ matrix:
\[ \mbox{\mcalR} = \left( \begin{array}{cc}
\partial e_1/\partial \gamma_1 & \partial e_2/\partial \gamma_1 \\
\partial e_1/\partial \gamma_2 & \partial e_2/\partial \gamma_2 \end{array} \right).\]
We can use the ensemble mean of such measurements \vest, for example, measured
from a population of galaxies, as a shear estimator.  Assuming the shear is
small, we can drop terms of order $\gamma^2$ and higher (we explore this
approximation in \S \ref{sec:weakshear}), such that
\begin{align}
    \langle \vest \rangle &= \langle \vest \rangle |_{\gamma=0} + \langle \mbox{\mcalR} \vecg \rangle + ... \nonumber \\
                          &\approx \langle \mbox{\mcalR} \vecg \rangle,
\end{align}
where we have also assumed the intrinsic ellipticities of galaxies are randomly
oriented such that $\langle \vest \rangle |_{\gamma=0} \sim (0,0)$.  If we have
estimates of \mcalR\ for each galaxy, we can form a weighted average:
\begin{align} \label{eq:rcorr}
    \langle \vecg \rangle &\approx \langle \mbox{\mcalR} \rangle^{-1}  \langle \vest \rangle \approx \langle \mbox{\mcalR} \rangle^{-1} \langle \mcalR \vecg \rangle.
\end{align}
Note the special case of constant shear, where $\gamma$ factors out of the
right-hand side.

If the estimator \vest\ is unbiased, the mean response matrix \mcalRmean\
will be consistent with the identity matrix.  If \vest\ is a biased
estimator, \mcalRmean\ will deviate from the identity matrix,
and could have significant structure.

The essence of \mcal\ \citep{HuffMcal} is to estimate the shear response
\mcalR\ for a measurement \vest\ directly from image data.  The measurement of
the shear estimator \vest\ is repeated on sheared versions of the galaxy
image and these are used to form a finite-difference central derivative.
For component $i,j$ we can write
\begin{equation} \label{eq:Rnum}
    R_{i,j} = \frac{\est_i^+ - \est_i^-}{\Delta \gamma_j},
\end{equation}
where $\est^+$ is the measurement made on an image sheared by $+\gamma$,
$\est^-$ is the measurement made on an image sheared by $-\gamma$,
and $\Delta \gamma = 2\gamma$.

The shearing is accomplished via a series of image manipulations. The original
image $I$ is deconvolved from the point spread function (PSF), sheared, and
reconvolved by the another function to suppress the amplified noise that is due to
deconvolution.  This function should be slightly larger than the original PSF
in order to suppress Fourier modes exposed by the shearing that were
previously hidden in the finite resolution of the original image.  We can
represent the series of operations clearly in Fourier space, where
convolutions are products and deconvolutions are divisions:
\begin{equation}
    \Itild(\gamma) = \left[ \left( \Itild/\Ptil \right) \oplus \gamma \right] \times \Ptild
\end{equation}
where \Itild\ and \Ptil\ are the Fourier transforms of the image and PSF.
\Ptild\ is the Fourier transform of the function with which the image
is reconvolved; we use the subscript $d$ to indicated that the function
is ``dilated'' with respect to the original PSF.
Shearing is represented by $\oplus$.

Note that for measurement of the shear estimator \vest, one should use an image
passed through these same image manipulations, but without any shear applied.
This ensures that the same reconvolution function is used for the shear
estimator and the response measurements.

We find that the results are rather insensitive to the choice of applied shear.
We tested values in the range of 0.001 to 0.05 and did not see significant
changes in the results. We take $\gamma=0.01$.

In practice, the response matrices measured from each image can have
significant structure, but the average \mcalRmean\ is to good approximation
diagonal.  Thus, for a straight mean shear measurement the correction in
equation \ref{eq:rcorr} reduces to element-wise division.  However,
measurements such as tangential shear or pairwise two-point functions involve
projecting the ellipticities into different coordinate systems. The measurement
may require projection of the response matrix as well, which could require
using the full response matrix.

One might expect the response to be proportional to the identity matrix if
there is no preferred direction in the measurement process.  In the simulation
tests we present in \S \ref{sec:sims}, we found that the diagonal elements can
differ by as much as a few parts in a thousand due to the use of a strong PSF
anisotropy oriented along the diagonal of the image.

As mentioned, the estimator \vest\ can be noisy, but in principle, when
averaging over a large ensemble of measurements, the noise does not cause any
bias because \mcalRmean\ is very well determined (see \S
\ref{sec:differences}).  However, when working with images, the \mcal\ process
itself alters the noise in a coherent way, requiring a correction (see \S
\ref{sec:fixnoise}).

\subsection{PSF Anisotropy and Shear Inference}  \label{sec:differences}

If the PSF correction is not perfect, the leading term $\langle \vest
\rangle|_{\gamma=0}$ will not be zero.  \cite{HuffMcal} discussed another
response, the response of the measurement to the PSF ellipticity, to correct
this effect.  In this work we instead reconvolve the image by a circular
function, which removes most additive effects that are due to the PSF  (see \S
\ref{sec:modelfit}).

In \cite{HuffMcal} the simple averaging in equation \ref{eq:rcorr} did not work
well because the shape estimators used  therein were relatively noisy.
Instead, a sophisticated statistical method was developed to infer the shear.
For the estimator we use in this work (see \S \ref{sec:modelfit}), the \mcalR\
are relatively well measured, even for galaxies with very low \snr, and the average
\mcalRmean\ is very well determined. We find that using simple averages in
equation \ref{eq:rcorr} is adequate to infer the correct shear.

\pagebreak
\section{Full Metacalibration Formalism} \label{sec:formalism}


In this section we derive the full \mcal\ formalism, including selection
effects. First we introduce some notation:  in principle, the estimator \vest\
can be any set of measurements from an image, but in what follows, without loss
of generality, we write the shear estimator \vest\ as a two-component
ellipticity $\vest = (\est_1,\est_2)$.  We write the ellipticity measured from
a sheared image as $\vest^+$ and $\vest^-$, for measurements on positive and
negatively sheared images, respectively.  We denote the selection probability as
$S$;  we can also make selections based on sheared parameters, which we denote
$S^+$ and $S^-$, the probability of selection after a positive or negative
shear is applied, respectively.

\subsection{Response for the Mean Shear} \label{sec:Rmean}

Suppose we wish to use the mean ellipticity as an estimator for the
mean shear.  The mean ellipticity over a large ensemble can be written as 
\begin{align}
    \langle \vest \rangle = \int P(\vest)~\vest~d\vest,
\end{align}
where $P(\vest)$ is the probability distribution of \vest.  We choose to work
with continuous functions so that all derivatives are well defined, in
particular the derivative of the selection function that we introduce below.

Assuming each galaxy experiences a small shear and that galaxy orientations
are random in the absence of shear, the mean ellipticity can be rewritten, to
leading order, as
\begin{align} \label{eq:basicR}
    \langle \vest \rangle \approx \int d\vest \frac{\partial P(\vest) \vest  }{\partial \vecg}\bigg|_{\gamma=0} \vecg ~ d\vest = \langle \mcalRg \vecg \rangle,
\end{align}
where we have ignored the perturbation of the normalization $\int d\vest P(\vest)$
because it leads to terms that are second order or higher in the shear.  The
mean shear is thus weighted by a response matrix \mcalRg.  This is the same $2
\times 2$ response matrix as discussed in \S \ref{sec:mcal}; we have added the
subscript $\gamma$ to differentiate this response from the selection response discussed
below.  If the \mcalRg\ are known, we can form a weighted-average estimator
for the mean shear:
\begin{align} \label{eq:full_rcorr}
    \left< \vecg \right> \approx \mcalRgmean^{-1} \left< \vest \right> \approx \mcalRgmean^{-1} \left< \mcalRg \vecg \right>.
\end{align}

We can calculate this mean response \mcalRgmean\ using quantities measured on
artifically sheared images, as discussed in \S \ref{sec:mcal}.  We will
approximate the derivatives using finite differences in the shear, such that
\begin{align} \label{eq:basicRfinite}
    \mbox{\mcalRgmean}  &= \int \frac{\partial P(\vest) \vest  }{\partial \vecg}\bigg|_{\gamma=0} d\vest
    \approx \int d\vest \left( \frac{ P^+ \vest_i^+ - P^- \vest_i^- }{\Delta \gamma_j}\right)  d\vest   \nonumber \\
    &= \frac{\langle \est_i^+ \rangle - \langle \est_i^- \rangle}{\Delta \gamma_j},
\end{align}
where we switched to component notation, such that
$i,j$ denotes the derivative of the $i$th ellipticity component with respect
to the $j$th shear component.  In practice, this averaging is performed
over an ensemble of measurements for discrete objects. It is equivalent to
averaging the responses as measured for each object.

\subsubsection{Selection Effects for the Mean Shear}

Now consider a selection that modifies the distribution of the measurement
\vest.  We will write this selection function as $S(\vest)$, the
probability of selecting an object with ellipticity \vest, although the
selection may be indirect, for example, a cut on \snr.  This selection function
could also represent some kind of weighting scheme that indirectly weights by
ellipticity.

After introducing a selection, the mean becomes
\begin{align}
    \langle \vest \rangle^S = \int S(\vest)~P(\vest)~\vest~d\vest.
\end{align}
We will assume the $\int d\vest P(\vest) S(\vest) = 1$, and continue to ignore
the higher order effect from changes in the normalization under shear.
Again, assuming a small shear and that galaxy orientations are random in the
absence of shear, the mean ellipticity after selection can be rewritten, to
leading order, as
\begin{align}
    \langle \vest \rangle \approx \int d\vest \frac{\partial S(\vest) P(\vest) \vest  }{\partial \vecg}\bigg|_{\gamma=0}~\vecg~ d\vest = \langle \mcalR \vecg \rangle,
\end{align}
Thus, the mean shear in the presence of selections is also weighted by a
response term \mcalR, and this response now includes the
shear response as well as the effects of the selections.
The probability that an object is selected changes after it is sheared.

This response with selections can be calculated using quantities measured on
artifically sheared images. It is useful to examine separately the response of
the estimator $\vest$ to a shear and the response of selection effects to a shear:
\begin{align}
    \mcalRmean  &= \int \frac{\partial S(\vest) P(\vest) \vest  }{\partial \vecg}\bigg|_{\gamma=0} d\vest \nonumber \\
    &= \int \left[ S(\vest) \frac{\partial P(\vest) \vest}{\partial \vecg}\bigg|_{\gamma=0} +  P(\vest) \vest  \frac{\partial S(\vest)}{\partial \vecg}\bigg|_{\gamma=0} \right] d\vest
\end{align}
Note that the first term is identical to the response in equation \ref{eq:basicR},
but now with selections applied.  As we show below, the second term represents the
response of selection effects to a shear.

We will again approximate the derivatives using finite differences in the shear.
Using the notation for measurements on sheared images, introduced
in \S \ref{sec:formalism}, we can rewrite the response as
\begin{align} \label{eq:Rmean}
    \mbox{\mcalRmean} &\approx
    \int d\vest \left [ S \left( \frac{ P^+ \est_i^+ - P^- \est_i^- }{\Delta \gamma_j}\right) + P \est_i \left( \frac{ S^+ - S^- }{\Delta \gamma}\right) \right] d\vest \nonumber \\
    &=\frac{\langle \est_i^+ \rangle^S - \langle \est_i^- \rangle^S}{\Delta \gamma_j} + 
\frac{\langle \est_i \rangle^{S+} - \langle \est_i \rangle^{S-}}{\Delta \gamma_j} \nonumber \\
       &\equiv \langle \mbox{\mcalRg} \rangle + \langle \mbox{\mcalRS} \rangle,
\end{align}
where $\langle \est^+ \rangle^S$ represents the mean of the ellipticity
measured from artificially sheared images, with selections based on parameters
measured on images without artificial shearing.  $\langle \vest \rangle^{S+}$
represents the mean ellipticity measured on images without artificial shearing, but with
selection based on parameters measured from artificially sheared images.

Thus the first term \mcalRgmean\ in equation \ref{eq:Rmean} is the average of
the shear responses measured for individual galaxies, the same as shown in
equation \ref{eq:basicRfinite}, but now with selections applied based on the
object parameters measured from images without artificial shearing.  The second
term \mcalRSmean\ represents the response of the selection effects to
a shear.



\setlist[enumerate]{leftmargin=5mm}

In order to calculate the desired weighted mean shear, one measures the
following:
\begin{enumerate}
	\item The mean ellipticity measured from unsheared images, selecting on
        measurements from unsheared images.   This is the mean shear estimator
		we wish to calibrate.
	\item The mean ellipticity measured from artificially sheared images, selecting
        on measurements from unsheared images,  $\langle e^+ \rangle^S$, $\langle e^- \rangle^S$. Alternatively, 
        one can form responses for each galaxy \mcalRg\ and average those.
    \item The mean ellipticity from unsheared images, selecting on measurements
        from positively and negatively sheared images $\langle e \rangle^{S+}$, $\langle e \rangle^{S-}$
\end{enumerate}



\subsection{Response for Two-point Functions} \label{sec:Rtwopt}

We can extend this formalism to two-point functions of the ellipticity also known
as shear-shear correlations.  In this type of measurement, the product of
ellipticites for objects or regions of sky separated by some finite distance is
averaged.  Because the mass causing the lensing effect is correlated, the
shears will be correlated as well. The two-point function can thus be interpreted
to constrain the correlations in the mass.

We will write the two-point function as
\begin{align}
    \xi &= \langle \vest_\alpha \vest_\beta \rangle \nonumber \\
        &= \int \mbox{\probe} \vest_\alpha \vest_\beta d \vest_\alpha d \vest_\beta
\end{align}
where the $\langle \vest_\alpha \vest_\beta \rangle$ indicates the mean
ellipticity product for pairs of objects at different locations, or two
separated regions of sky.  Including selection effects, this becomes
\begin{align}
    \xi &= \int \mbox{\probe} S_\alpha S_\beta \vest_\alpha \vest_\beta d \vest_\alpha d \vest_\beta
\end{align}
where the selections are independent for each object.  Note that we have used
the shorthand $S(\vest_\alpha) \rightarrow S_\alpha$.

Adopting the symmetry assumptions used in \S \ref{sec:Rmean} and additionally
assuming that the mean shear is zero over a large area of sky used to measure
the two-point function, the leading term in the Taylor expansion is a second
derivative,
\begin{align} \label{eq:basictwopt}
\xi &\approx \int d \vest_\alpha  d\vest_\beta  \frac{\partial^2 S_\alpha S_\beta \mbox{\probe} \vest_\alpha \vest_\beta}{\partial \vecg_\alpha \partial \vecg_\beta}\bigg|_{\gamma=0}  \vecg_\alpha \vecg_\beta \nonumber \\
    &\equiv \left<  \mbox{\mcalRtwopt} \vecg_\alpha \vecg_\beta  \right>.
\end{align}
The next largest terms are of order $ \langle e_\alpha \rangle|_{\gamma=0} \gamma_\alpha^2$.
In equation \ref{eq:basictwopt} the
true correlation function has been weighted by a response \mcalRtwopt. The mean of this response
is given by
\begin{align} \label{eq:Rtwoptfull}
    \mbox{\mcalRtwoptmean}  = 
    \int d \vest_\alpha  d\vest_\beta  \frac{\partial^2 S_\alpha S_\beta \mbox{\probe} \vest_\alpha \vest_\beta}{\partial \vecg_\alpha \partial \vecg_\beta}\bigg|_{\gamma=0}.
\end{align}
If we know the response \mcalRtwopt, we can form an estimator for the correlation function
that takes the form of a weighted average:
\begin{align} \label{eq:twoptavg}
    \xi \approx  \mcalRtwoptmean^{-1} \langle \vest_\alpha \vest_\beta \rangle \approx \mcalRtwoptmean^{-1} \left<  \mbox{\mcalRtwopt} \vecg_\alpha \vecg_\beta  \right>.
\end{align}

Note that the joint probability distribution \probe\ is not separable 
because the shear at each galaxy $\alpha$ is correlated with that at galaxy
$\beta$. However, assuming the shapes of galaxies are not correlated in the absence of
lensing, \probe\ is separable at zero shear such that $\mbox{\probe}|_{\gamma=0}
= P(\vest_\alpha)|_{\gamma=0} P(\vest_\beta)|_{\gamma=0}$.  The following identities
also hold:
\begin{align} \label{eq:identities}
    \frac{\partial \mbox{\probe}}{\partial \vecg_\alpha}\bigg|_{\gamma=0} &= P(\vest_\beta) \frac{\partial P(\vest_\alpha) }{\partial \vest_\alpha} \frac{\partial \vest_\alpha}{\partial \vecg_\alpha}\bigg|_{\gamma=0} \nonumber \\
    \frac{\partial^2 \mbox{\probe}}{\partial \vecg_\alpha \partial \vecg_\beta}\bigg|_{\gamma=0} &= \frac{\partial P(\vest_\alpha) }{\partial \vest_\alpha} \frac{\partial P(\vest_\beta) }{\partial \vest_\beta}  \frac{\partial \vest_\alpha}{\partial \vecg_\alpha} \frac{\partial \vest_\beta}{\partial \vecg_\beta}\bigg|_{\gamma=0}.
\end{align}
It then follows that the response can be completely factored such that
\begin{align} \label{eq:Rseparable}
    \int d & \vest_\alpha  d\vest_\beta  \frac{\partial^2 S_\alpha S_\beta \mbox{\probe} \vest_\alpha \vest_\beta}{\partial \vecg_\alpha \partial \vecg_\beta}\bigg|_{\gamma=0}  \nonumber \\
      & \approx \int d \vest_\alpha  \frac{\partial S_\alpha P(\vest_\alpha) \vest_\alpha}{\partial \vecg_\alpha}\bigg|_{\gamma=0} \int d\vest_\beta   \frac{\partial S_\beta P(\vest_\beta) \vest_\beta}{\partial \vecg_\beta}\bigg|_{\gamma=0} \nonumber \\
      &=  \mcalRmeanalpha \mcalRmeanbeta.
\end{align}

For cross-correlations, the separate mean responses given in equation
\ref{eq:Rseparable} are different and should be calculated as indicated.
For auto-correlations, these two responses \mcalRmeanalpha\ and
\mcalRmeanbeta\ are identical, and are each equal to the
response of the mean shear given in equation \ref{eq:Rmean}.  In
that case, the response
of the two-point auto-correlation function is approximately given by
\begin{align}
    \left< R^{2pt} \right> \approx \mbox{\mcalRmean}^2.
\end{align}

Thus it may be sufficient when measuring two-point functions to calculate the
mean response for the ensemble of objects.  This is a reflection of the
assumption that \probe\ is separable at zero shear.  However, this
assumption may not hold in general if spatially correlated objects have
correlated shapes in the absence of lensing (see \S \ref{sec:IA}) or if
imperfections in image reduction and object identification cause
biases that depend on the local galaxy density (see \S
\ref{sec:summary}).


In what follows we test the formulas for mean shear.  We will test the
response for two-point functions in a future work.

\subsection{Remarks on Selections}

Selections that produce additive effects are not addressed in the above.  If a
circular reconvolution function used, as we will do (see \S
\ref{sec:modelfit}), then selections on object properties will not be
correlated with the PSF shape, by construction, and thus no net additive bias
will be introduced. Note, however, that preselections, for example detection
thresholds, will in general produce such biases.  We will test the importance
of preselections with image simulations below.

It is important that the selections be placed well above any detection
thresholds, such that after shearing, detected objects can move into and out
of the selected sample.  If the selection were too close to a detection
threshold, some objects that should become detectable after
a shear would not be included.

In real data, a selection may directly or indirectly select the redshift of the
sources, which means that the true shear will be different after selection,
even in the absence of selection biases.  This must be taken into account
separately.  We do not treat this effect in what follows.

Note that if the selection function $S(\vest)$ is a step function, the
derivative of $S(\vest)$ in \ref{eq:Rmean} will be infinite.  Thus one should
avoid placing threshold cuts directly on the ellipticities.  However, it is
valid to place cuts on other observables, such as \snr, which will result in a
smooth selection on the ellipticity.

\subsection{Remarks on Intrinsic Alignments} \label{sec:IA}

In the above we assumed that \probe\ is separable at zero shear, but this
assumption breaks down for physically associated galaxies
\citep[e.g.][]{HirataIntrinsicAlign07}, an affect known as ``intrinsic
alignments'' \citep[IA; for a recent review, see][]{TroxelIAReview2015}.
Within the \mcal\ formalism, the non-separable nature of \probe\ can be partly
dealt with by calculating the full pair-weighted response in equation
\ref{eq:Rtwoptfull}, without the approximations leading to equation
\ref{eq:Rseparable}.  However, the signal itself will be biased by the presence
of IA.  A number of methods  have been proposed to correct for the
contaminating effect of IA in the shear \citep{TroxelIAReview2015}.  With
\mcal, we can include the correct shear weighting that occurs as a result of
the non-unity response of the shear estimate, which should improve the accuracy
of the corrections (see \S \ref{sec:weighting}).

\section{Contamination of the Response by Correlated Noise} \label{sec:contam}

In the presence of noise, the observed image can be written $I_o=I+\eta$, where $\eta$
is the ``noise image.''  The \mcal\ sheared images $I_o(\gamma)$ will contain
contributions from deconvolved, sheared and reconvolved noise. Again working
in Fourier space, and using the notation introduced in \S \ref{sec:mcal},
\begin{align}
    \Itild_o(\gamma) &= \left[ \left( \Itild/\Ptil + \ntil/\Ptil \right) \oplus \gamma \right] \times \Ptild  \nonumber \\
    &= \Itild(\gamma) + \ntil(\gamma).
\end{align}

The deconvolution correlates the noise across the image.  This correlated
noise is sheared and then reconvolved by \Ptild, producing the
sheared correlated noise image $\eta(\gamma)$.  The pattern of this sheared
noise is coherent between the positively sheared and negatively sheared images.
The effect is small, but it is amplified through division by $\Delta \gamma$ to
form the central derivative.  We thus expect the observed response \mcalRo\ to
be contaminated by the response of the correlated sheared noise which
we denote \mcalRnoise:
\begin{equation}
    \mbox{\mcalRo}  =  \mcalR + \mbox{\mcalRnoise}.
\end{equation}
For the simulations and fitting method employed in this work, we generally
found \mcalRnoise\ to be $-5$ to $-10$\%.


\subsection{Correction for Sheared Correlated Noise} \label{sec:fixnoise}

We explored four different empirical corrections.  We describe our favored
method here; the others are described in appendix \ref{sec:altcorr}.

This method, which we refer to as ``\fixnoise'', is designed to statistically
cancel the effects of correlated noise caused by the \mcal\ procedure.  For each
image that was passed through the convolution and shearing steps, producing
an image $I(\gamma)_o$, we generated a random noise field
$\eta_r$, and applied the same operations, but using a shear with
the opposite sign:
\begin{equation}
    \widetilde{\eta}_r(-\gamma) = \left[ \left( \ntil/\Ptil \right) \oplus (-\gamma) \right] \times \Ptild  \nonumber \\
\end{equation}
We then added this image in real space to the $I(\gamma)_o$ image
\begin{equation}
    \hat{I}(\gamma) = I_o(\gamma) + \eta_r(-\gamma).
\end{equation}
The result, $\hat{I}(\gamma)$, is an approximation for the image
without correlated noise.
We used these $\hat{I}(\gamma)$ to measure the shapes and responses used in
shear recovery.  

This procedure necessarily increases the noise in the measurements and shear
recovery.  We test the increased noise in \S \ref{sec:degrade}.

Because the noise is increased, it is important that both the response and
estimator should be measured on the $\hat{I}(\gamma)$ images, so that the
response is representative of the correct noise level.

Note that we have assumed that the original noise in the image is uncorrelated.
This assumption does not hold in coadded images because of the interpolation
that must occur to place all images in the same coordinate system.  In order to
apply this correction, it is therefore simplest to work with the original
images rather than a coadded image.  For a description of such a processing
scheme, see, for example, \cite{Jarvis2016}.


\section{Weighted Averages for Associated Parameters} \label{sec:weighting}

As discussed in \S \ref{sec:formalism}, if a shear estimator has
non-unity response, the mean shear (or two-point function) is effectively
weighted.  If we know the responses, we can form a weighted average.

When averaging associated quantities, such as redshifts, one should also weight
by these responses.  For example, a quantity $x$ should be averaged as
\begin{align}
    \left< x \right> = \frac{\sum_i R_i x_i}{\sum R_i}.
\end{align}
The responses are not scalar, but it is probably sufficient
to use the average of the two diagonal components of the $R$ matrix
for this purpose. One concern is that the measured $R$ values are not
strictly positive.  It is worth exploring whether this causes
any bias in real scenarios.

\section{Image Simulations} \label{sec:sims}

We applied \mcal\ to a set of ``postage stamp'' image simulations.  We used two
different types of simulations, one based on parametric galaxies and another
based on real galaxy images.  Information about each simulation type is give in
table \ref{tab:sims}.  In both simulations, only a single object was present
in the image to avoid the effects of blending.

\begin{table*}
    \centering

    \begin{tabular}{ l c c c c c c c c}
		\tableline
        \rule{0pt}{3ex}Sim          & Galaxy Model      & Size \& Flux   & PSF Model   & PSF FWHM        & PSF shape     & Shear & \# Galaxies    & \# Stars     \\
                    &                   &             &             & [arcsec]        &               &       &               &             \\
		\tableline
		\tableline
        \rule{0pt}{3ex}\rgsim       & \cosmosname       & \cosmosname & Kolm./Opt.  & \psfrdist/Opt.  & Optical       &  Variable & \nsimNgal     & None            \\
        \bdksim      & \bdkfull  & \galrdist   & Moffat      & \psfrdist       & \nsimPSFShape &  \nsimShear & \nsimNgalBD & None            \\
        \bdstar      & \bdkfull  & \galrdist   & Moffat      & \psfrdist       & \nsimPSFShape &  \nsimShear & \nsimNgalBD & \nsimNStar       \\
		\tableline
    \end{tabular}

	\parbox{0.9\textwidth}{
		\caption{\begin{flushleft}Description of the image simulations.  For the \bdksim\
		simulations, the Bulge and Disk had independent ellipticities but the same
		half light radius \hlr.  The galaxy \hlr\ and flux were drawn from fits to the 25.2 mag.
		limited COSMOS sample.  Additionally, knots of ``star formation'' were added as
		point sources distributed as a random walk in the disk.  The \bdstar\
		simulation shared the same galaxy images with the \bdksim\ simulation, but
		with additional star images included.  For
		the \rgsim\ simulation, real COSMOS images were used for the galaxies. For
		the PSF, a  Kolmogorov atmospheric turbulence model was used, 
		plus contributions from an optical model matched to DES; the mean PSF size
		was approximately 0.9 arcsec. The PSF ellipticity is given in the ``reduced
		shear'' convention.  For comparision, the typical DES PSF ellipticity is about
                0.01 in these units.
		\label{tab:sims}
		\end{flushleft}}
	}
\end{table*}

\subsection{Simulations with Parametric Models} \label{sec:bdksim}

We created an image simulation using complex parametric models.  We modeled
galaxies as a bulge and disk, with additional knots of star formation.  We
label this simulation \bdkfull\ (\bdksim).

We used different uncorrelated ellipticities for the bulge and disk
components, but gave both bulge and disk the same half-light radius \hlr.
We drew the \hlr\ and flux from fits \citep{LacknerGunn2012} to the 25.2
magnitude limited sample from the COSMOS data
\citep{Scoville2007a,Scoville2007b}, as distributed with the \galsim\
simulation package \citep{GALSIM2015}.  In order to avoid repeating the exact
parameters, we interpolated the joint \hlr-flux distribution using a kernel
density estimate.

The ellipticities were drawn from the simple model used in \cite{bfd2016},
\begin{align} \label{eq:edist}
    P(e) &\propto \left[1-(e)^2\right]^2 \exp\left[-e^2/2\sigma^2\right],
\end{align}
with $\sigma=0.2$ for the disk and $\sigma=0.1$ for the bulge.  Note that we
used the ``reduced-shear'' style ellipticity
\begin{equation}
    e = \frac{1-q}{1+q},
\end{equation}
where $q$ is the axis ratio \citep{bj02}.  This definition is roughly a
factor of two smaller than the ``distortion'' style ellipticity
used in \citet{bfd2016}.

For the simulated knots of star formation, we distributed 100 point sources
according to a random walk starting at the disk centroid, with each point
taking 40 steps.  This random walk method was inspired by the simulation
presented in \cite{Zhang2008FourierQuadI}\footnote{Our implementation of the
random walk galaxy is available as a \galsim\ object class
\texttt{galsim.RandomWalk}}.  We then distorted this distribution using the
same ellipticity assigned to the disk; but note that this does not result in a
purely elliptical distribution of point sources.  Only in the rare case of a
pure bulge or disk with no knots is the model purely elliptical.

The fraction of the flux in the bulge was drawn uniformly between zero and
unity.  The fraction of the disk flux assigned to knots of star formation
was then also chosen uniformly between zero and unity.  The resulting 
objects range from pure bulge, bulge+disk, pure disks, bulge+disk+knots, 
to pure knots. A model with pure knots resembles an irregular galaxy.

Finally, the entire composite \bdkfull\ object was sheared with the same
shear, $\vecg = (0.02, 0.00)$.

We convolved the galaxy images with a PSF modeled as a Moffat profile
\citep{Moffat1969}, with $\beta=3.5$ and FWHM=0.9 arcsec. The PSF ellipticity
was set to $e_2 = 0.025$ in reduced-shear units ($\sim0.05$ in units of
distortion). 

These convolved objects were randomly offset from the image center by $\pm 0.5$
pixels, and rendered onto a 48 by 48 grid, with pixel scale 0.263 arcsec per
pixel. Constant Gaussian noise was added, with the same noise level used for
all images. 

\begin{figure*}[p]
    \centering
    \includegraphics[]{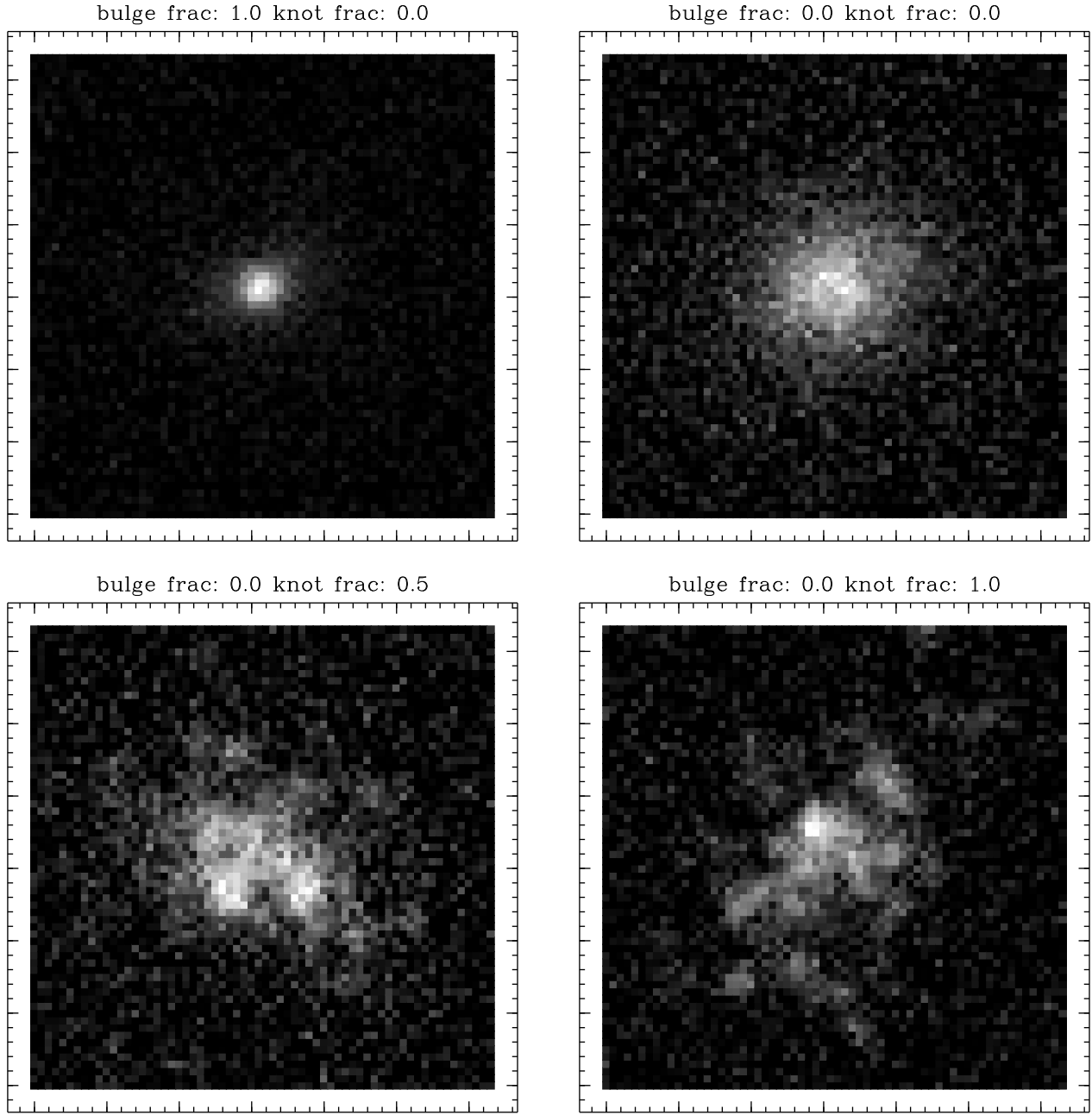}

    \caption{Example of simulated galaxy images.  Each image is a composite of a
    bulge and disk, plus knots of star formation.  The half-light radius is the
    same for all components, while the fraction of light in each component
    varies.  In the upper left and upper right we show pure bulge and pure disk
    models, respectively.  In the lower left we show a disk with half the light
    in knots, and in the lower right we show a pure ``irregular'' galaxy
    composed entirely of knots.  Each model was convolved by a Moffat PSF
	and pixelized.  For demonstration purposes, we here show very
    large models to make the detailed structure visible; the galaxies used for
    our shear tests are typically much smaller than the PSF (see figure
	\ref{fig:psimhlrcompare}). }

	\label{fig:parametricgals}

\end{figure*}

In figure \ref{fig:parametricgals} we show some example simulated galaxies with
a range of parameters.  Here we have shown large models, much larger than the
PSF, in order to show detailed internal structure.

\begin{figure}
    \centering
    \includegraphics[width=\columnwidth]{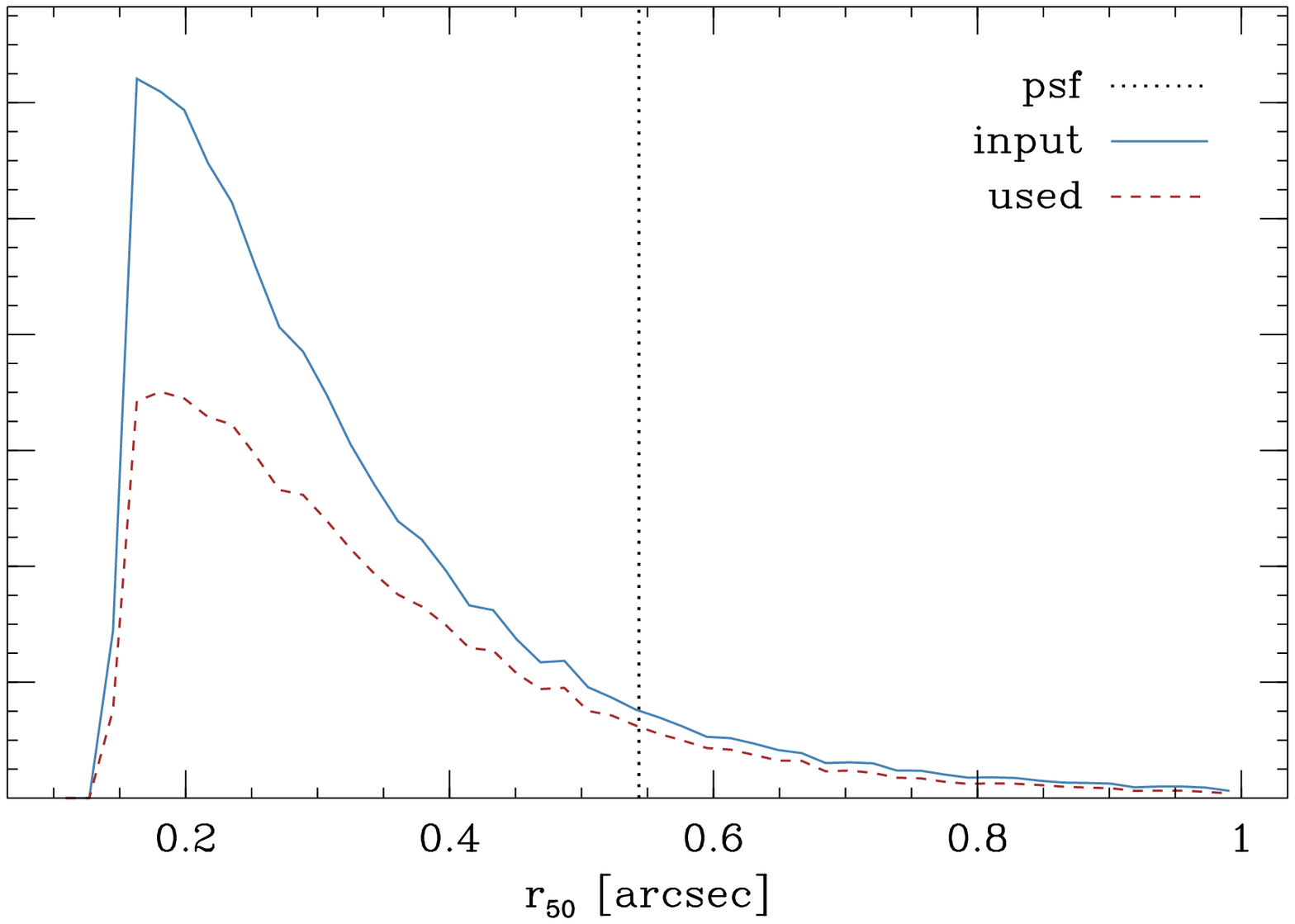}

    \caption{Distribution of half-light-radius \hlr\ in the parametric simulations.
        The solid line represents the distribution of input \hlr, drawn from fits
        to COSMOS data.  The dashed line represents the \hlr\ for objects that passed
		the initial $S/N > 5$ pre-cut.  The \hlr\ of the PSF is shown as the vertical dotted
        line.}

\label{fig:psimhlrcompare}
\end{figure}

In figure \ref{fig:psimhlrcompare} we show the distribution of \hlr\ for the
parametric galaxies, along with the \hlr\ for the PSF.  Note that most objects
had \hlr\ significantly smaller than the PSF.

\begin{figure}
    \centering
    \includegraphics[width=\columnwidth]{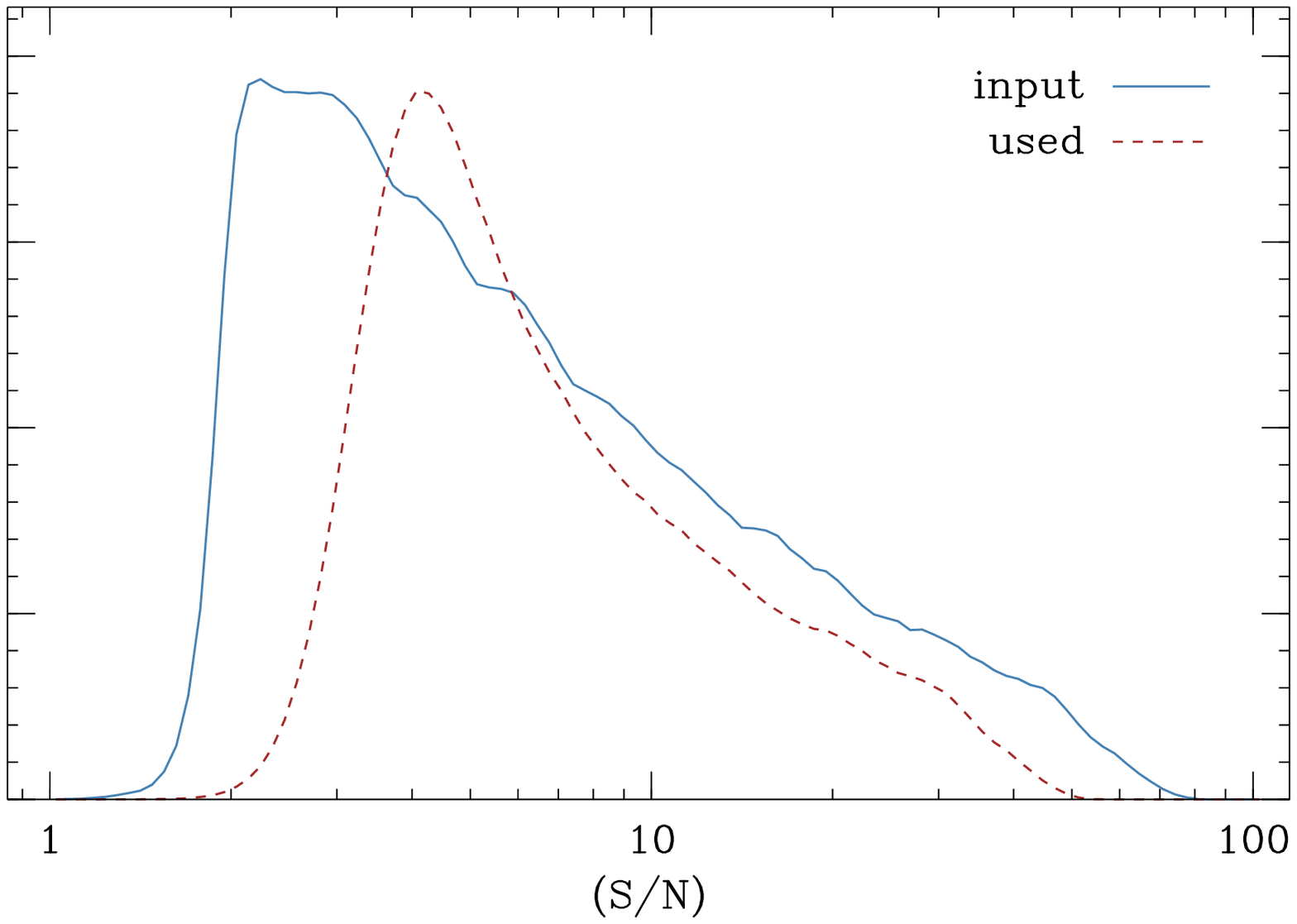}

    \caption{Distribution of S/N in the parametric simulations. The
    solid curve represents the true input distribution, the dashed curve 
	represents the objects that passed the initial pre-cut on {\it measured}
    \snr$ > 5$. The measured \snr\ was biased and noisy,
    resulting in a smooth selection on true \snr.  This pre-cut
    does not sharply cut on magnitude and results in a catalog that is
    limited at COSMOS i-band magnitude $\sim$25.}

\label{fig:s2n}
\end{figure}

In figure \ref{fig:s2n} we show the distribution of \snr\ for the parametric
galaxies.   Shown are both the true input \snr\ and the distribution of true
\snr\ after applying a cut at {\it measured} \snr$ > 5$.  This ``true'' \snr\
is that calculated by \galsim, and is the maximal \snr\ based on the true model
\citep{Jarvis2016}.  The measured \snr\ is noisy and biased (see \S
\ref{sec:results}), being derived from a single-Gaussian model, so the cut
results in a smooth rolloff in the true \snr.  This \snr\ preselection results
in a noisy cut on magnitude.  The resulting input catalog for \mcal\ has a
limiting magnitude of COSMOS i-band $\sim$25, but with a significant number
of objects removed at brighter magnitudes.

\subsubsection{Simulated Stars}

In order to test the robustness of \mcal\ to stellar contamination, we included
$\sim$\nsimNstarperc\ stars in the \bdksim\ simulation.  The stars were simply
drawn as PSFs, with the same flux distribution as used for the galaxies.  We
refer to the simulation with stars as \bdstar; the galaxies are identical to
those in the \bdksim\ simulation, but stars were then included in the analysis.

\subsubsection{Preselection} \label{sec:preselect}

In real data, detections with significance less than \snr$\sim 5$ can be
spurious, so in practice a threshold must be placed to produce what are
considered reliable detections.  Such a preselection can produce significant
selection biases.  A round object has a higher \snr\ than if it were sheared, and
galaxies oriented in the same direction as the PSF have a higher \snr\ than
objects otherwise oriented.  Thus a preselection will tend to alter the
distribution of ellipticities, which will bias the shear recovery.  In
particular, if this selection occurs before \mcal\ and object fitting, the
corrections for selection effects presented in \S \ref{sec:formalism} cannot be
used.

In the \bdksim\ simulation, we generated images with \snr\ as low as $\sim 2$,
as shown in figure \ref{fig:s2n}.  In order to test the effect of a
preselection, we did not perform the \mcal\ process on all images.  We applied
a preselection on objects with {\em measured} \snr$ > 5$, and these objects
were removed from the analysis.  Then, as discussed in \S
\ref{sec:shear_recover}, we applied further cuts on the measured \snr\ above
this threshold in order to minimize the bias that is due to removal of these galaxies.

\pagebreak
\subsection{Real-Galaxy Simulations} \label{sec:cosmosim}

We designed a second set of simulations to mimic the ``real-galaxy'' constant
shear simulations used in the GREAT3 challenge \citep{great3}, which used
real galaxy images from COSMOS data.  The galaxies were
drawn from the 23.5 magnitude limited sample distributed
with the \galsim\ simulation package \citep{GALSIM2015}.  We generated 1000
different fields in which galaxies were given a constant shear ranging from
0.01 to 0.08, with random orientations.

We implemented two important changes as compared to GREAT3.  First, we oriented
the galaxies randomly, whereas in GREAT3 the galaxies were placed in pairs
rotated by 90 degrees, in order to cancel shape noise.  Using paired galaxies has
the undesired effect of cancelling some biases that we wish to explore
\citep{Jarvis2016}.  Second, we used optical aberrations in the PSF designed to
match that seen in the Dark Energy Survey data\footnote{Aaron Roodman, private
communication}.  Similar to GREAT3, we varied the aberrations as Gaussian
random variables around a fiducial value. These root-mean-squared variations,
in units of waves in the Noll convention \citep{noll1976}, are given in table \ref{tab:aberr}.
We used a Kolmogorov model for the atmospheric component, such that
the overall mean FWHM $\sim 0.9$ arcsec for 0.263 arcsec pixels.
Each galaxy was rendered onto a 48 by 48 pixel grid.
For this configuration there are
significant variations in the PSF ellipticity, but relatively little net
ellipticity across the entire simulation.  The code used to
generate these simulations began as a fork of the GREAT3 public code base, and
is freely available online\footnote{\url{https://github.com/esheldon/egret}}.

\begin{table}
    \centering
    \begin{tabular}{  l  l  }
		\tableline
        \rule{0pt}{2ex} Zernike Component  & RMS Variation \\
        \tableline
        \tableline
        \rule{0pt}{2ex}Defocus & 0.13 \\
        Astigmatism in Y & 0.13 \\
        Astigmatism in X & 0.14 \\
        Coma in Y & 0.06 \\
        Coma in X & 0.06 \\
        Trefoil in Y & 0.05 \\
        Trefoil in X & 0.06 \\
        Spherical & 0.03 \\
		\tableline
    \end{tabular}

	\parbox{0.8\columnwidth}{
		\caption{\begin{flushleft}Root-mean-squared variation for the aberrations in the optical model,
			in units of waves in the Noll convention, derived 
		from Dark Energy Survey data. \label{tab:aberr}\end{flushleft}}
	}
\end{table}

\begin{figure*}
    \centering
    \includegraphics[width=0.8\textwidth]{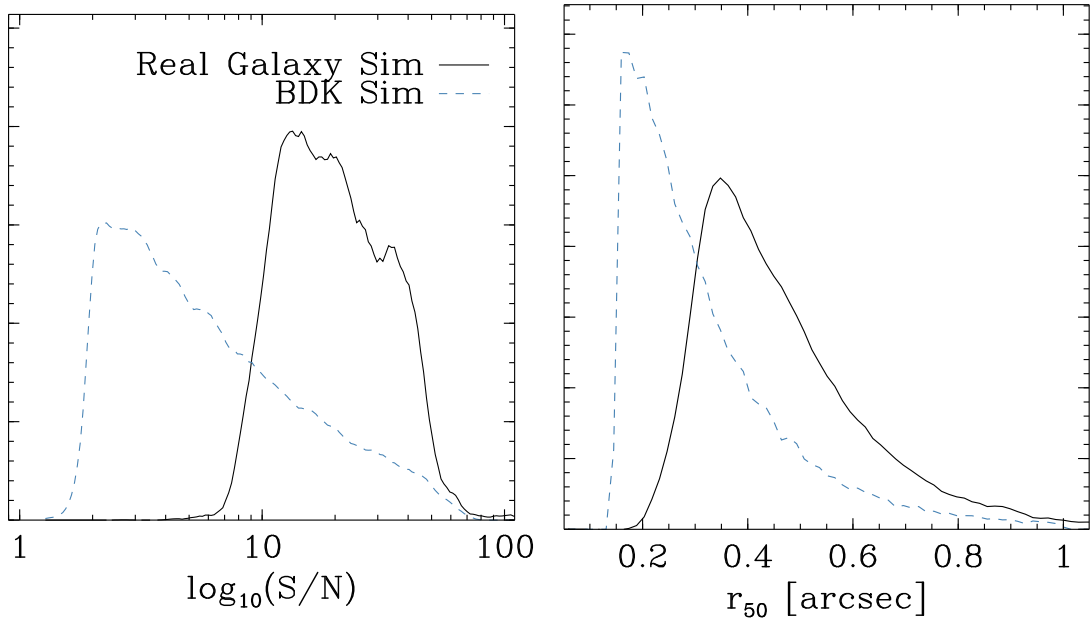}

		\caption{Distribution of properties in the COSMOS real galaxy simulations. The
		left panel contains the distribution of measured S/N, while the right panel contains
		the distribution of half-light-radius from the cosmos catalog for the input
		galaxies.  For comparison, the distributions for the \bdkfull\ \bdksim\ simulations
		are overplotted as dashed lines.}
\label{fig:cosmos}
\end{figure*}

In figure \ref{fig:cosmos} we show the distribution of measured \snr\ for the
COSMOS simulations, as well as for the \bdkfull\ simulation
\bdksim.  Also shown is the distribution of the half-light-radius \hlr\
for the two simulations.

Note that the galaxies used in the parametric sims presented in \S
\ref{sec:bdksim} are much fainter and smaller than those used in the \rgsim\
simulation.  Also note that the \rgsim\ simulation was relatively expensive
compared to the parametric simulations, so we generated fewer galaxies and did
not implement any preselection.  We thus do not expect the real galaxy
simulation to be more challenging than the parametric simulation in every
aspect. We include it to directly test the robustness of \mcal\ to special
properties of real galaxies that may not appear in our \bdkfull\ simulation,
and to test shear recovery using a more realistic PSF.  

\section{Model Fitting and Metacalibration Operations} \label{sec:modelfit}

We fit the images with a single Gaussian model using the \ngmix\
code\footnote{\url{https://github.com/esheldon/ngmix}}.  To perform the fit we
used an implementation of the ``adaptive moments (AM)'' algorithm originally
presented in \cite{bj02}.   We applied no PSF correction.  We expected this
estimator to respond weakly to a shear, exhibit large model bias, noise bias,
and bias due to lack of PSF correction.  Note that we also applied \mcal\ to a
forward-modeling, maximum likelihood estimator; we give some brief results for
that method in appendix \ref{app:maxlike}.

In order to correct for PSF anisotropy, we reconvolved by a symmetrized version
of the PSF. We created this PSF by adding the PSF image to itself, rotated by
90, 120, and 180 degrees.  This averaging can result in a Fourier space image
that is larger in some dimensions than the original, so we further shrunk
the symmetrized PSF in Fourier space.  The shrink factor was taken to be
$1+2*\delta$, where
\begin{align}
    \delta = \frac{E}{T/2}
\end{align}
Here $E$ is the maximum eigenvalue from the covariance matrix of the best-fit
Gaussian. This we divide by half the trace $T$, which is the mean extent of the
object.  For a purely elliptical PSF, a factor of $1+\delta$ would be
sufficient; we conservatively increase the factor to $1+2\delta$ in case the
Gaussian fit does not completely capture the asymmetries of the true PSF.

This symmetrization method requires that an image of the PSF is available at
the location of the object, which is already a requirement of \mcal\ in order
to perform deconvolutions.  To produce such an image, one must accurately model
stars and interpolate to the location of each object, for example as provided
by the \psfex\ package \citep{BertinPSFEx2011}.  Note that instead of using a
round reconvolution function, one may instead use the response of the estimator
to a PSF shear to address any uncorrected affects of PSF anisotropy
\citep{HuffMcal}.

All \mcal\ image operations were performed using the \texttt{metacal} module
from \ngmix, which in turn uses \galsim\ to perform most image manipulations.
We used the correction for correlated noise, as discussed in \S
\ref{sec:fixnoise}

\section{Results} \label{sec:results}

In what follows, we will characterize the bias using the standard linear model
\citep[e.g.][]{great3} with a multiplicative part $m$ and an additive part $c$,
such that
\begin{align}
	\left< \gamma_{meas} \right> = (1+m) \gamma_{true} + c.
\end{align}
For the \bdksim\ simulations, there was only one shear value (\nsimShear),
and the PSF had ellipticity only in one component (0.000, 0.025). We
thus determined $m$ from the first component and $c$ from the
second.  For the \rgsim\ simulations, there were many different shears,
so the linear model above was fit.  We found the multiplicative
bias was the same in each component, so we combined them into
a single value in the plots and tables below.

In all cases the measured \snr\ is based on the best-fit Gaussian model (true
parameters for the simulated galaxies, such as scale radius or \snr, were never
used in the analysis).  We used a similar definition to that implemented in the
GREAT3 simulations \citep[][equation 16]{great3}, but replaced the true profile
with the model
\begin{equation}
    (S/N)^2 = \frac{1}{\sigma^2} \int m(x,y)^2 dx dy,
\end{equation}
where $m$ is the value of the model at location $x,y$, and $\sigma^2$ is the
variance of the noise in the image. A Gaussian is a poor fit in general, so
this measure of the \snr\ is biased and noisy.

\subsection{Metacalibration Responses}

\begin{figure}[h]
    \centering
    \includegraphics[width=\columnwidth]{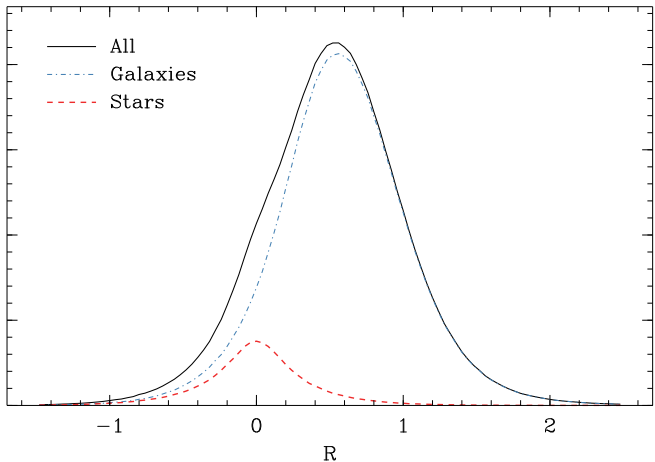}

    \caption{Distribution of \mcal\ responses for galaxies and stars in
    the \bdstar\ simulation.  Stars have mean response close to zero,
    and thus do not bias the overall shear calibration.}

\label{fig:Rstars}
\end{figure}

In figure \ref{fig:Rstars} we show the measured \mcal\ responses for the
\bdksim\  simulations.  Also shown is the response with the stars included in
the \bdstar\ simulation.  The distribution of \mcalR\ is quite symmetric in the
absence of stellar contamination.  We will discuss the affect of stars in \S
\ref{sec:stars}.

\subsection{Shear Recovery} \label{sec:shear_recover}

In table \ref{tab:results} we show results for shear recovery in each of our
simulations.  As was discussed in \S \ref{sec:preselect}, we applied a
preselection to the \bdksim\ simulation at measured \snr$ > 5$, which imposes
a selection bias.  We thus placed cuts at higher \snr\ than this threshold, so
that the corrections for selection effects presented in \S \ref{sec:formalism}
could be used accurately.  In this table we show results for \snr$ > 10$.  For
the \rgsim\ simulation we did not apply a preselection.

\begin{table}
    \centering
    \begin{tabular}{ l  ccc}
		\tableline
        \rule{0pt}{3ex}Sim    & $m$               & $c_1$            & $c_2$ \\
               & $[10^{-3}]$       & $[10^{-5}]$      & $[10^{-5}]$ \\
        \tableline
		\tableline
        \rule{0pt}{3ex}\rgsim & $0.22 \pm 0.58$  & $2.6 \pm 2.9$    & $2.2 \pm 2.9$ \\
        \bdksim & $0.03 \pm 0.31$  & -                & $-0.15 \pm 0.62$ \\
        \bdstar& $0.01 \pm 0.32$  & -                & $-0.08 \pm 0.63$ \\
		\tableline
    \end{tabular}

	\parbox{0.9\columnwidth}{
		\caption{\begin{flushleft}\Mcal\ results for each image simulation described in
			\S \ref{sec:sims} and table \ref{tab:sims}.  For each simulation,
			a cut was placed at \snr$ > 10$, and corrections were applied
			for selection effects (see table \ref{tab:results_sel} for more
			results on selections).  A single Gaussian was fit
		to the observed object, with no PSF correction applied.
		No multiplicative or additive bias was detected in any case. 
		Stellar contamination at the level of \nsimNstarperc\ increases
		the noise in the recovered shear by \starnoiseincrease\ but does not introduce 
		a significant bias.  
		\label{tab:results}\end{flushleft}}
	}

\end{table}

Using \mcal\ we found no significant multiplicative or additive biases.
Without applying the \mcal\ responses, the multiplicative bias $m$ was of order
50\% for all simulations.


\subsubsection{Results with Selection Effects}

\begin{figure}[h]
    \centering
    \includegraphics[width=\columnwidth]{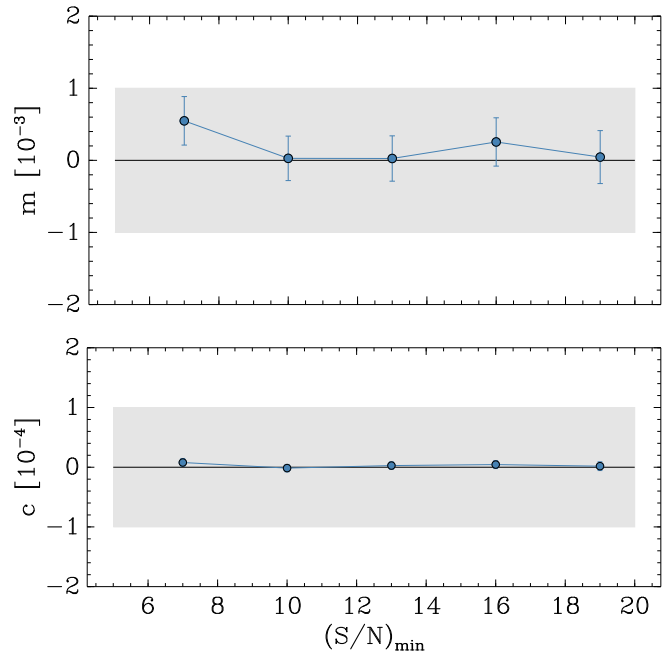}

    \caption{Multiplicative (upper panel) and
		additive bias (lower panel) in the \bdksim\ simulation after applying
        threshold selections in the measured signal-to-noise ratio \snr.   
        The filled gray region represents the target accuracy. } 

\label{fig:s2nthresh}
\end{figure}

\begin{figure}[h]
    \centering
    \includegraphics[width=\columnwidth]{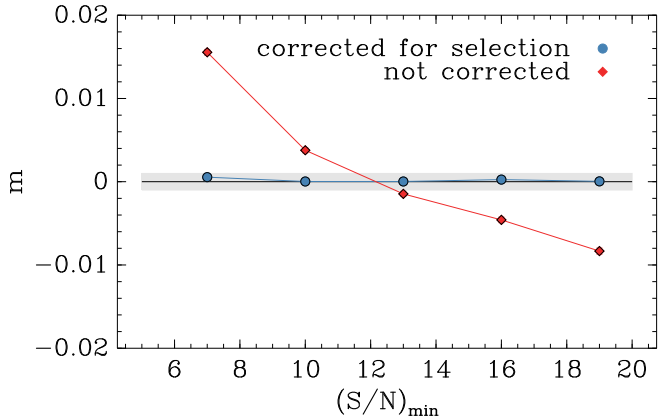}

	\caption{Same as the top panel of figure \ref{fig:s2nthresh}, but now
additionally showing the multiplicative bias without corrections for selection
effects.  The bias without correction for selection effects is represented as
red diamonds. The bias after correction for selection effects is represented as
blue circles.  The filled gray region represents the target accuracy. } 

\label{fig:s2nthresh_nocorr}
\end{figure}

In table \ref{tab:results_sel} we show the results for different \snr\
threshold cuts in the \bdksim\ simulations.  We show the recovered bias with and
without corrections for selection effects.  These results are also shown
graphically in figures \ref{fig:s2nthresh} and \ref{fig:s2nthresh_nocorr}.  The
cuts were all placed above the preselection at \snr$ > 5$ to guarantee the
validity of the corrections.

\begin{table*}
    \centering
    \begin{tabular}{ l cc  cc}
        \tableline
       \rule{0pt}{3ex} & \multicolumn{2}{c}{Uncorrected for Selection}                      & \multicolumn{2}{c}{Corrected for Selection} \\
        Selection                   & $m$             & $c$            & $m$               & $c$  \\
                                    & $[10^{-3}]$     & $[10^{-5}]$    & $[10^{-3}]$       & $[10^{-5}]$ \\
        \tableline
        \tableline
\rule{0pt}{3ex}$\mbox{\snr} > 7 $ & $+15.55 \pm 0.34$ & $+0.80 \pm 0.68$ & $+0.55 \pm 0.34$ & $+0.79 \pm 0.67$ \\
$\mbox{\snr} > 10 $ & $+3.78 \pm 0.31$ & $-0.15 \pm 0.62$ & $+0.03 \pm 0.31$ & $-0.15 \pm 0.62$ \\
$\mbox{\snr} > 13 $ & $-1.46 \pm 0.31$ & $+0.27 \pm 0.63$ & $+0.03 \pm 0.31$ & $+0.27 \pm 0.63$ \\
$\mbox{\snr} > 16 $ & $-4.58 \pm 0.33$ & $+0.44 \pm 0.67$ & $+0.26 \pm 0.34$ & $+0.44 \pm 0.67$ \\
$\mbox{\snr} > 19 $ & $-8.33 \pm 0.36$ & $+0.17 \pm 0.73$ & $+0.05 \pm 0.37$ & $+0.18 \pm 0.73$ \\
        \tableline
    \end{tabular}

	\parbox{0.9\textwidth}{
		\caption{\begin{flushleft}\Mcal\ results for the \bdksim\ simulation with various
			cuts on signal-to-noise ratio \snr.   Results are shown with and without corrections
			for selection effects.
		\label{tab:results_sel}\end{flushleft}}
	}
\end{table*}

We measured and corrected for a significant multiplicative selection bias in
each case.  These biases are generally well above our desired
part-in-a-thousand accuracy.  After correction, we found that the
multiplicative bias was less than a part in a thousand in all cases.  We did
not find any additive selection biases, which suggests that our procedure of
reconvolving by a symmetrized PSF was sufficient for these simulations.

\subsubsection{Results with Stellar Contamination} \label{sec:stars}

Results including \nsimNstarperc\ stars in the \bdstar\ simulation are shown in
table \ref{tab:results}. We did not detect any additional bias after including
stars. The noise in the recovered shear did, however, increase by
\starnoiseincrease.

\Mcal\ is robust to stellar contamination if the PSF is well characterized.
Images consistent with a PSF will not, in the mean, respond to the shear
applied during the \mcal\ process.  Measurement on stars also yields zero
average shape as long as the PSF correction is sufficiently accurate: our use
of a symmetrized PSF (see \S \ref{sec:modelfit}) appears to be sufficient in
this case.  In figure \ref{fig:Rstars} we show the measured response \mcalR\
for stars and galaxies.  Indeed we see that for stars, the \mcalR\ is noisy
but consistent with zero.  Thus, in the mean, stars contribute zero to both the
estimator and response, leaving equation \ref{eq:rcorr} unbiased.

If the additional variance is tolerable, it may be useful to include stars in a
shear analysis if the PSF is sufficiently well known.  Attempting to remove
faint stars from a sample is a noisy procedure, likely to induce selection
effects.  These can be controlled using the corrections derived in \S
\ref{sec:formalism}, but only if the selection is also repeated based on
quantities measured on sheared images, so the corrections can be calculated.
If the selection must be performed outside of the \mcal\ process, it may be
better to avoid it altogether.

For accurate interpretation of the signal, it is important to weight by the
\mcal\ response terms in order to obtain the correct redshift distribution (see \S
\ref{sec:weighting} for more discussion of weighted means).  It is also
desirable that the redshift estimates for stars be close to zero, so that the
weighted redshift distribution is minimally contaminated.

\subsubsection{Effects of Missing Data}

The Fourier transforms used to perform the \mcal\ convolutions cannot
accommodate missing data.  In real data there are features in the image,
however, such as bad pixels and columns, and cosmic rays that cannot be used
for object measurement.  This can be dealt with easily when the galaxy model is
fit simultaneously to postage stamps drawn from all available observing epochs
and bands \citep[e.g.][]{Jarvis2016}.  If the epochs are spatially offset, such
that the object does not always appear at the same location in the image, then
a small fraction of images should be affected by missing data such as bad
pixels.  In that case, data deemed problematic can simply be left out of the
fit.

If only a single image is available, one may wish to replace the missing data.
To test this scenario, we chose to replace the missing data with the value from
the best fit Gaussian model.  Using a better model would be preferred; this may
be considered a worst-case scenario.

We ran a simulation in which 10\% of the images had bad pixels or bad columns.
We found that the model replacement worked well for single bad pixels. For bad
columns we found a large additive $e_1$ bias, as well as a multiplicative bias
of a few parts in a thousand.  However, this bias disappeared if we introduced
a compensatory ``bad column'' at 90 degree rotation about the center of the
image, restoring symmetry to the image.   Again, we wish to emphasize that such
a procedure may not be necessary when there are many images available for
fitting.

\subsubsection{Noise Degredation Due to Correlated Noise Corrections} \label{sec:degrade}

As a result of the noise added to correct for correlated noise (see \S
\ref{sec:fixnoise}), the \snr\ of the measurements after the \mcal\ procedure
are reduced by $\sqrt{2}$, but this not a severe limitation.  For faint
galaxies, which dominate the sample, the shape measurement noise is comparable
to the intrinsic shape noise, so one might expect the increase in effective
shear noise to be less than $\sqrt{2}$.

We performed a test where the mean shear was calculated in a simulation, with
the correlated noise correction.  This was compared to a ``perfect'' \mcal\
procedure without correlated noise. To accomplish this, we added noise after
the \mcal\ image manipulations were performed, rather than before.  In both
cases we placed a cut at $\snr\ > 10$. For the simulations described in \S
\ref{sec:bdksim} and the fitting used in \S \ref{sec:modelfit}, we found that
the uncertainty in the shear recovery was increased by $\sim$\degrade.  Note
that one may be tempted to decrease the \snr\ cut to recover objects that have
a lower \snr\ after the additional noise is added, but we find that including
objects with \snr\ $< 10$ actually increases the variance.  This can also be
seen in the results shown in table \ref{tab:results_sel}.  We found this to be
true even when inverse variance weighting was introduced.

For correlation function measurements over large scales, such as shear-shear
correlations, sample-variance will dominate and this \degrade\ increase will be
relatively unimportant.  Nevertheless, we consider it worthwhile to explore
alternative corrections that do not involve adding significant noise.

\section{Weak Shear Approximation} \label{sec:weakshear}

In deriving the response terms in \S \ref{sec:formalism}, we assumed the shear
was small, so that the response of the estimator to a shear was linear. This
approximation will break down at higher shears.  We expect
the multiplicative bias to have the following generic form \citep{bfd2016}:
\begin{align} \label{eq:breakdown}
	B(\gamma) = m + \alpha \gamma^2 + O(\gamma^4)
\end{align}

We tested the bias from nonlinearity using both the \bdksim\ image
simulations and a ``toy model,'' inspired by that used in \citep{ba14}.
For the \bdksim\ image simulations, we ran tests with shears of
0.06 and 0.10, in addition to the 0.02 discussed above.

For the toy model we drew ellipticities from the distribution given in equation
\ref{eq:edist} and added constant shear analytically using the standard shear
transformation equations \citep{SeitzSchneider97}.  We then multiplied by a
bias factor of 0.6 and added Guassian noise to each ellipticity component with
scatter of 0.2. We truncated the total ellipticity to be less than unity,
making the noise effectively non-Gaussian.  We performed \mcal\ operations by
analytically shearing the shapes.  Note that we did not shear the noise, so
sheared noise effects as seen in the image simulations are not present.

\begin{figure}
	\centering
    \includegraphics[width=\columnwidth]{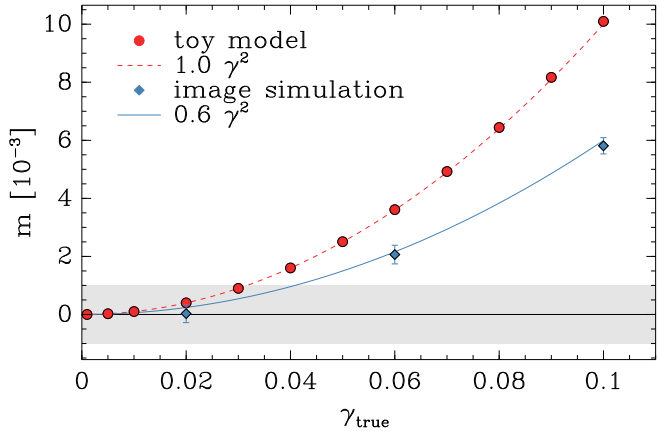}

    \caption{Multiplicative bias as a function of applied shear in two types of
    simulations.  The circles represent the results for the toy simulation
    presented in the text.  The diamonds represent results for the parametric
    \bdksim\ image simulations.  For the toy simulation, the bias scales with
    $\gamma^2$.  For the image simulation the bias scales as $0.6 \gamma^2$.
    The gray region represents the target accuracy.}

\label{fig:weaklens}
\end{figure}

In figure \ref{fig:weaklens} we show the results.  Fitting the bias model in
equation \ref{eq:breakdown}, we found $m \sim 0$ and $\alpha \sim 1$ for the
toy model.  For the image simulations we found $m \sim 0$ and $\alpha \sim
0.6$. We note that \cite{bfd2016} also found that the value of $\alpha$
depended on the type of simulation used, with values for $\alpha$ as high as
2.  Taking the value of $\alpha$ from the image simulation, we find that the
nonlinearity in this simulation becomes greater than our part-in-a-thousand
goal for shears greater than about 0.04.  Using $\alpha=1$ from the toy model,
we reach the threshold for shears higher than about 0.03.

Following the discussion in \cite{bfd2016}, we expect the bias on a cosmic
shear measurement to be approximately $1 + 3\alpha\sigma_\gamma^2$, with a
shear variance of $\sigma_\gamma^2 \sim 0.02^2$.  For the $\alpha=0.6$ found in
the image simulations, the bias would be about $7.2 \times 10^{-4}$. 
For $\alpha=1$, the bias would be be about $1.2 \times 10^{-3}$, exceeding our
desired accuracy and requiring some correction.  As pointed out by
\cite{bfd2016}, this correction does not require great precision.  If the bias
were determined at the $\sim 40$\% level for $\alpha=1$,  the shear could be
recovered accurately with 95\% confidence.  More care may be required for
measurements of higher shear, such as tangential shear measurements near the
centers of galaxy clusters.  We will explore strategies to mitigate this
potential bias in a future work.  

\section{Computation Time}

\mcal\ involves performing image manipulations five times, one for the zero
shear, but reconvolved image, and four for each of the sheared images.
We then perform image fitting on each of these images.

The total computation time per object was approximately 0.2 seconds.  In our
tests the time was dominated by the image manipulations, primarily the Fourier
transforms used for convolutions.  The image manipulations dominated partly
because we used the fast \ngmix\ code; fitting each image took only about 0.001
seconds, so about 0.005 to fit all five images.  For slower fitting codes, the
\mcal\ manipulations may be subdominant.

It may be that further optimization is possible.  For example, a code that
works entirely in Fourier space could avoid rendering the reconvolved image in
real space.

\section{Summary and Future Work} \label{sec:summary}

We introduced a formalism for \mcal\ to calculate shear response corrections,
including corrections for selection effects. We developed a simple empirical
correction for the correlated noise associated with the \mcal\ procedure.

We tested the formalism using simulations based on real galaxy images, as well
as challenging parametric simulations that included preselection effects and
stellar contamination.  We applied a range of cuts on the object \snr, inducing
significant selection affects.  In all cases we recovered the input shear to
better than a part in a thousand.

\Mcal\ compares favorably with other shear measurement techniques.  At the time
of writing, the only technique with demonstrated accuracy close to \mcal,
without reliance on calibration from simulations, and which accurately
addresses selection effects, is \bfd\ \citep{bfd2016}.  \mcal\ has been tested
using more challenging simulations and has proved more accurate.  Unlike \bfd,
\Mcal\ does not rely on significant prior information about galaxy properties.
On the other hand, in our current implementation we add extra noise to deal
with correlated noise effects.  This extra noise increases the uncertainty in
the shear recovery by about \degrade\ in our tests. The relative increase in
noise should be smaller for studies that are sample-variance limited, such as
shear-shear correlations on large scales. Nevertheless, we have identified the
development of a more precise correlated noise correction as a priority going
forward.

In order to reach part-in-a-thousand accuracy in real data, there are a number
of additional challenges to be addressed.  These challenges are shared with
other shear measurement techniques.  We think the most important are the
effects of overlapping objects and the limitations of current image processing
techniques. The accurate determination of the PSF is also fundamental, but we
do not address that here.

Biases in the image processing pipeline that identifies image artifacts,
determines the background light, identifies unique detections, and assigns
light to objects will potentially produce biases in the shear recovery.  Image
artifacts may not be properly corrected for, or masked.  In crowded regions
such as galaxy clusters, the pipeline may fail to determine the background
accurately.  The \mcal\ process may not accurately determine the shear response
in these cases.

In crowded regions the detection algorithm may fail to identify all the unique
objects, even those brighter than the detection threshold, because objects
overlap significantly on the sky.  The \Mcal\ procedure will correctly estimate
the shear response for blended objects, as long as the presence of neighbors
does not result in an instability in the measurement process.  This is
sufficient for blends at the same redshift, but for blends at different
redshift, this combined shear response has no simple interpretation because each
object will experience a different shear.  Ideally, we would identify the unique
objects and assign a fraction of the light in each pixel to each object.  Then
separate shear responses would be measured for each object, and redshifts of
the objects would be determined, at least statistically.

There will also be large numbers of faint galaxies that are not bright enough
to exceed the detection threshold. This is an important source of bias when
using simulations to calibrate the shear \citep[see e.g.][]{Hoekstra2017}, but
we have not yet tested how this will affect \mcal.  These objects will be
included at some level in the \mcal\ shear response measured for brighter
galaxies, and will contaminate the light used for flux and color determination.
The concern in this case is the same as for the bright blends mentioned above:
unique objects must be identified and the redshift distribution determined in
order to properly interpret the shear measurement.  The presence of
unidentified neighbors will complicate this process.   Going forward, it is
imperative to assess the importance of this effect on both the shear response
measurement and redshift determination.  

It may be possible to address these issues by inserting artificial galaxy
images into the data using something like the \balrog\ framework
\citep{Balrog2016}, followed by rerunning processing pipelines, including
shear estimation.  \balrog\ uses the original images as the basis, so all the
features of real observations need not be simulated.  The approach is
particularly applicable to studying the effects of blending, undetected
objects, and image artifacts.

The formalism we have developed recovers a weighted average of the shear, and
this weighting is known, being the same responses used to calibrate the shear
estimate.  In this work we only tested the simplest case of constant shear.
When using \mcal\ in real data, where the shear and response vary across the
sky, it will be important to accurately propagate this weighting, for example
when inferring the redshift distribution of a source population.

\section*{Acknowledgments}

ES is supported by DOE grant DE-AC02-98CH10886.

Many thanks to Paul Stankus for helpful comments on the draft.We are grateful
to Mike Jarvis and Rachel Mandelbaum for many useful discussions and help with
the \galsim\ package. Thanks to Gary Bernstein for suggesting our response
formula for two-point functions could be significantly simplified.  We thank
Aaron Roodman for providing the variation of the optical aberrations measured
in DES data.  Thanks to Matt Becker for help using the egret simulation
package.  We thank the anonymous referee for helpful comments and questions
that led to a significantly improved text.

\appendix

\section{Alternative Methods for Correlated Noise Correction} \label{sec:altcorr}

\subsection{Correction using \galsim\ Methods}

With guidance from the \galsim\ developers, we attempted to use the \galsim\
noise isotropization and whitening functionality to correct the correlated
noise.  Isotropization enforces four-fold symmetry, introducing minimal extra
noise, while whitening completely whitens the image, introducing significant
extra noise.  However, neither of these methods improved the shear recovery in
our simulations.  It may be that some aspects of the \mcal\ procedure
invalidate the assumptions behind these correction methods.

\subsection{Detrending the Correlated Noise Bias} \label{sec:detrend}

\subsubsection{Expected Scaling of the Bias with Noise Level} \label{sec:scaling}

The bias in the ellipticity that is due to correlated noise should scale with
the noise correlation function, and thus the square of the noise level in the
image $n^2$ \citep{Kaiser2000,HirataCorrNoise}.  As discussed in \S
\ref{sec:contam}, this bias will propagate into the response.  We can write the
observed \mcalRo\ as a contribution from both the actual response \mcalR\ and a
noise term \mcalRnoise\
\begin{align} \label{eq:scaling}
    \mcalRo &= \mcalR + \mcalRnoise  \nonumber \\
            &= \mcalR + A n^2.
\end{align}

\subsubsection{Detrending Correction Scheme}

We add a small
amount of noise to the image such that $n \rightarrow n + \Delta n$.  If we
then run the new image through the \mcal\ process, we can measure
$R^{\mathrm{before}}$, a response that will include correlated noise effects.
We can write this observed response as
\begin{align}\label{eq:Rbefore}
    \mcalRo^{\mathrm{before}} &= \mcalR + A (n + \Delta n)^2 \nonumber \\
       &\simeq \mcalR + A n^2 + 2 A n \Delta n
\end{align}
where we have dropped terms of order $(\Delta n)^2$ and higher.  In equation
\ref{eq:Rbefore}, \mcalR\ is the response at noise $n+\Delta n$ in the absence
of correlated noise.  

We can also add identical noise {\em after} the original image  has been run
through \mcal, and measure $R^{\mathrm{after}}$.  The response when adding
noise after \mcal\ does not suffer any additional bias due to correlated noise:
\begin{align}
    \mcalRo^{\mathrm{after}} &= \mcalR + A n^2.
\end{align}
The difference between these responses is then 
\begin{align}
    \Delta \mcalR &\equiv \mcalRo^{\mathrm{before}} - \mcalRo^{\mathrm{after}}  \nonumber \\
             &\simeq 2 A n \Delta n.
\end{align}

We propose the following procedure to correct for correlated noise:
\begin{enumerate}
    \item Calculate $\Delta \mcalR$ for a series of noise offsets $\Delta n$.
    \item Average $\Delta \mcalR$ over all objects for each noise offset.
    \item Perform a linear fit to $\Delta \mcalR$ vs. $2 n \Delta n$ to find the 
        coefficient $A$.
    \item Apply a mean correction for correlated noise given by
        \begin{align}
            \mcalRnoise & \simeq A n^2.
        \end{align}
\end{enumerate}
If the noise varies between observations, we can apply a 
correction based on the mean variance $A
\langle n^2 \rangle$.

\subsubsection{Measurements of the \detrend\ Parameters}

\begin{figure}
	\centering
    \includegraphics[width=\columnwidth]{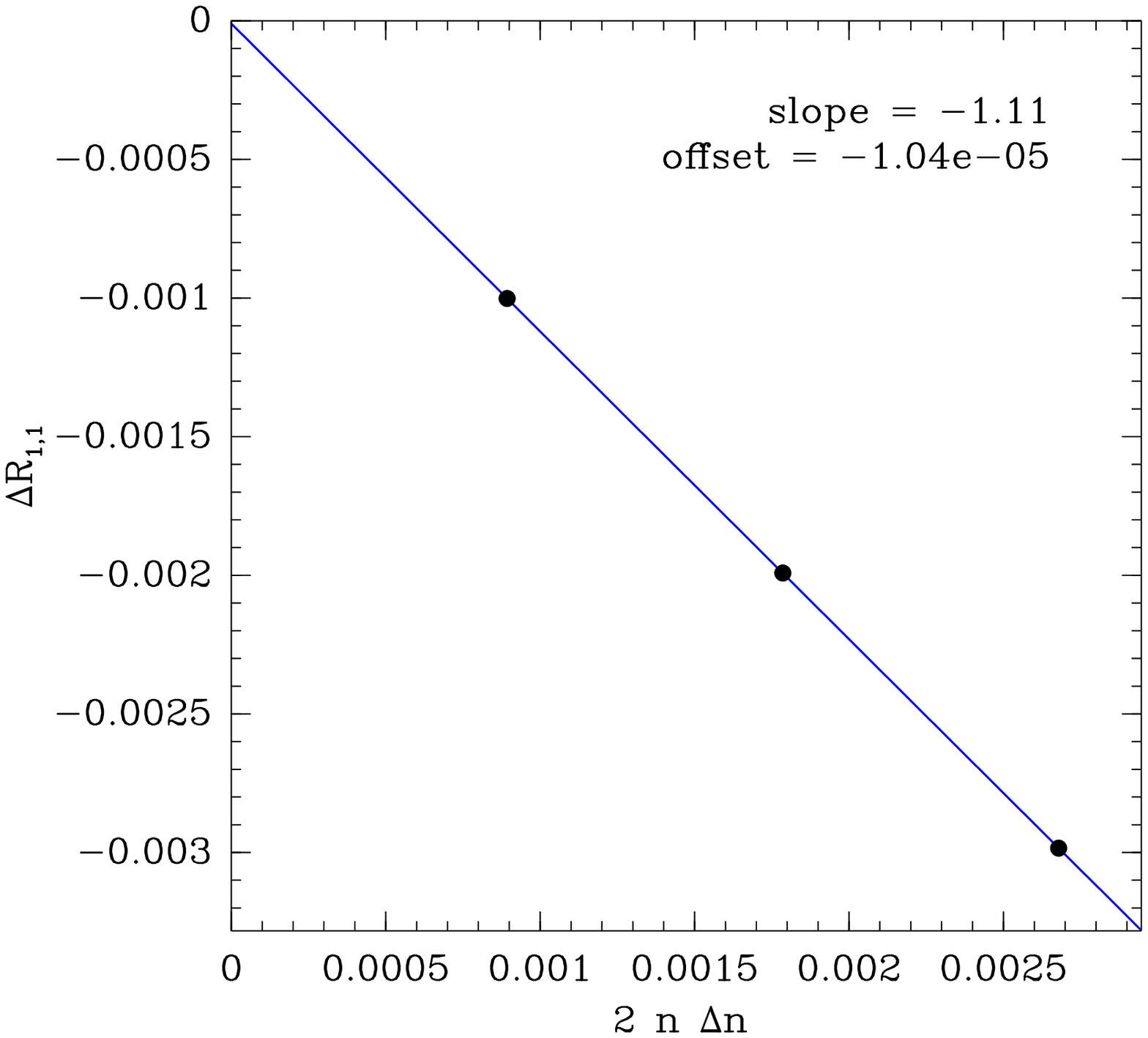}

    \caption{Trend of $\Delta R_{1,1}$ with $2 n \Delta n$ for the
        real galaxy simulation.   $n$ is the
    original noise level and $\Delta n$ is the additional noise added.  The
    trend is linear as predicted.}

\label{fig:detrend}
\end{figure}

We measured $\Delta R$ vs $2 n \Delta n$ to find the
coefficient $A$.  In figure \ref{fig:detrend}, we show this fit for the \rgsim\
simulation.  The trend is well fit by a linear model, as expected, with a slope
$A \simeq $\Aslope, implying a correction $A n^2 \simeq$ \Rcorr\ for this
simulation.

\subsubsection{Using a Random Subset To Calculate \Detrend\ Corrections}

Measuring the \detrend\ parameters requires extra computations, at least a
factor of three to fit the linear parameters given in \S \ref{sec:detrend}, and
a factor of four if an additional point is used to check for possible
nonlinearity. These extra computations could be expensive for large surveys.  

However, the \detrend\ parameters are more precisely measured than the shear
itself.  Here we explore the precision of the recovered shear using smaller
subsets of the object catalog to measure the \detrend\ parameters, and
applying the corrections to the full sample.  In Table \ref{tab:subsets}
we show the shear recovery parameters for various subset sizes for the
\bdksim\ simulation described in \S \ref{sec:results}

We find that a relatively small sample can be used to determine the correlated
noise correction.  Calculating the detrending terms for 10\% of the sample
leads to only 0.3\% extra variance in the recovered shear, and using 1\% of
galaxies leads to only 4.4\% increase in variance.

To calculate these numbers, we have assumed the extra uncertainty is added
quadratically with the uncertainty measured using all galaxies to estimate the
\detrend\ parameters; e.g for the first row, we have added approximately 30\%
quadratically with the measured uncertainty, resulting in a net increase of
4.4\%.

It is important to use a truly random subset of the population to determine the
corrections, including a fair sample of stars and other contaminants, and a
representative amount of pixel level masking.  If a particular aggregate shear
measurement involves a selection, this selection must also be applied to the
random subset.

\subsubsection{Performance of the \detrend\ method}

We found that this method did not work as well as the \fixnoise\ method
described in \S \ref{sec:fixnoise}.  We detected a remaining bias of $m \sim 2
\times 10^{-3}$ in the \rgsim\ simulations.

\subsection{Simulating Models}

In this method, we generated model images with the correct noise level
corresponding to each real image.  We then measured the response of the noise
that is due to the convolutions and shears used in \mcal, with noise added
before and after the \mcal\ procedure.

The measurement with correlated noise will be the sum of the response
without correlated noise plus the response of the correlated noise field,
\begin{equation}
    \mcalRnoisemodel = \mcalRmodel + \mcalRnoise.
\end{equation}
This measurement is quite noisy for a single galaxy, but we
can estimate the mean correlated noise response for an ensemble
of galaxies,
\begin{equation}
    \langle \mcalRnoise \rangle = \langle \mcalRnoisemodel \rangle - \langle \mcalRmodel \rangle.
\end{equation}
Each entry used in this average corresponds to the best-fit model
and noise properties for a galaxy in the sample.

The response \mcalRnoise\ can be subtracted to recover an estimate of the mean
response without correlated noise,
\begin{equation}
    \langle \mcalR \rangle = \langle \mcalRo \rangle - \langle \mcalRnoise \rangle.
\end{equation}

We detected significant bias using both exponential and Gaussian
models for the galaxy.  We found a remaining bias of  $m \sim 4 \times
10^{-3}$ in the \rgsim\ simulations.

\begin{table}
    \centering
    \begin{tabular}{ c  c }
        Subset Size & Extra Error \\
        \hline
        1\% & 4.4\% \\
        5\% & 0.8\% \\
        10\% & 0.3\% \\
    \end{tabular}

    \parbox{0.7\columnwidth}{
		\caption{\begin{flushleft}Additional variance in the recovered shear 
			using differently sized subsets to
			estimate the detrending corrections.  Values were obtained
		from 100 bootstrap samples. \label{tab:subsets}\end{flushleft}}
	}

\end{table}

\section{Metacalibration Using Maximum-likelihood Forward Modeling} \label{app:maxlike}

In \S \ref{sec:results} we discussed in detail the results for a moment based
fitting method using adaptive moments (AM), with no explicit PSF correction.
We also performed tests using a forward-modeling maximum likelihood method.
In this method we fit a single Gaussian to the PSF and then fit a Gaussian to each
object, convolved analytically by the Gaussian PSF model.  This method should
be more sensitive because the PSF is accounted for in the modeling, although
imperfectly.  We applied smooth priors to the likelihood for each parameter in
the model in order to ensure a stable fit, but these priors were not tuned to
match the true parameters of the simulated galaxies.

\begin{figure}[h]
	\centering
    \includegraphics[width=\columnwidth]{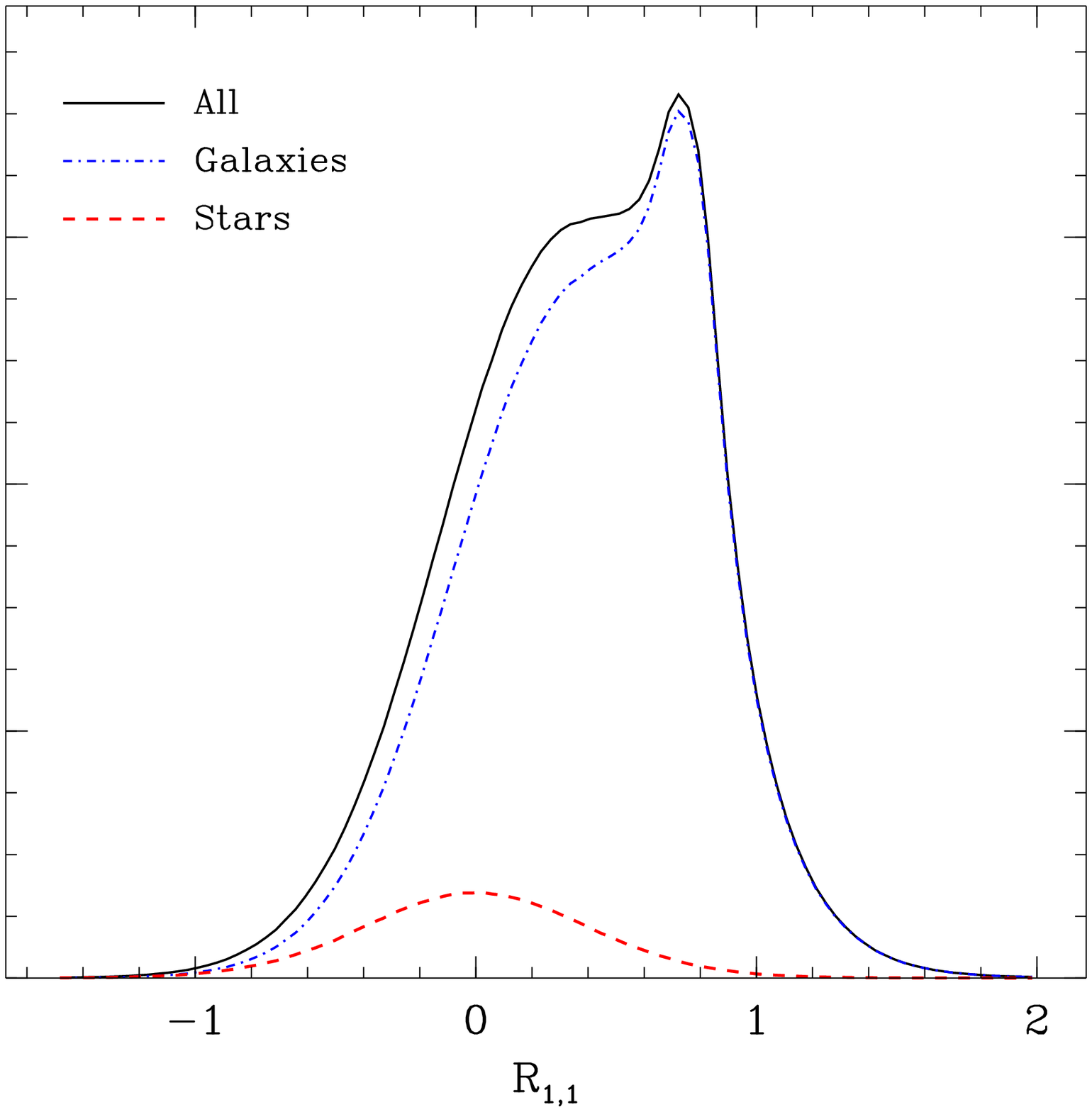}

    \caption{Distribution of shear response in the \bdkfull\ simulations
    for a shear estimator based on  forward-modeling maximum likelihood fitting.
    The colors and line styles are the same as that used 
    in figure \ref{fig:Rstars}.}

\label{fig:Rmaxlike}
\end{figure}

In the \bdkfull\ simulation, we found this method to perform equally as well as
AM, but with some interesting differences.  The method is indeed more
sensitive: as shown in figure \ref{fig:Rmaxlike}, the distribution of the
response has a narrow peak near 0.7, corresponding to disk-dominated galaxies
with a high \snr, with a tail to low response from a mix galaxies with low
\snr.  Note also that the expected value for perfect response for this
estimator is unity, whereas for AM it would be approximately twice as large
because the estimator for AM is a distortion style ellipticity rather than a
reduced shear style ellipticity.  However,  we do not see correspondingly large
decrease in the variance of the shear estimator, so it is not clear whether
this increased sensitivity is necessarily an advantage.

We also note that the distribution of response for the forward-modeling
estimator is less symmetric than for AM without PSF correction (see figure
\ref{fig:Rstars}), but this did not bias the shear recovery.  We see that the
distribution of response for stars is still quite symmetric about zero for this
estimator.

It was noted in \cite{HuffMcal} that for some very noisy estimators, the
resulting complex distribution of response can be problematic.  For this
reason, we suggest that practitioners explore the performance of their chosen
estimator in simulations before applying it to real data.

\bibliographystyle{mnras}
\bibliography{apj-jour,astroref}

\begin{thebibliography}{}
\makeatletter
\relax
\def\mn@urlcharsother{\let\do\@makeother \do\$\do\&\do\#\do\^\do\_\do\%\do\~}
\def\mn@doi{\begingroup\mn@urlcharsother \@ifnextchar [ {\mn@doi@}
  {\mn@doi@[]}}
\def\mn@doi@[#1]#2{\def\@tempa{#1}\ifx\@tempa\@empty \href
  {http://dx.doi.org/#2} {doi:#2}\else \href {http://dx.doi.org/#2} {#1}\fi
  \endgroup}
\def\mn@eprint#1#2{\mn@eprint@#1:#2::\@nil}
\def\mn@eprint@arXiv#1{\href {http://arxiv.org/abs/#1} {{\tt arXiv:#1}}}
\def\mn@eprint@dblp#1{\href {http://dblp.uni-trier.de/rec/bibtex/#1.xml}
  {dblp:#1}}
\def\mn@eprint@#1:#2:#3:#4\@nil{\def\@tempa {#1}\def\@tempb {#2}\def\@tempc
  {#3}\ifx \@tempc \@empty \let \@tempc \@tempb \let \@tempb \@tempa \fi \ifx
  \@tempb \@empty \def\@tempb {arXiv}\fi \@ifundefined
  {mn@eprint@\@tempb}{\@tempb:\@tempc}{\expandafter \expandafter \csname
  mn@eprint@\@tempb\endcsname \expandafter{\@tempc}}}

\bibitem[\protect\citeauthoryear{{Bernstein}}{{Bernstein}}{2010}]{Bernstein2010}
{Bernstein} G.~M.,  2010, \mn@doi [\mnras] {10.1111/j.1365-2966.2010.16883.x},
  \href {http://adsabs.harvard.edu/abs/2010MNRAS.406.2793B} {406, 2793}

\bibitem[\protect\citeauthoryear{{Bernstein} \& {Armstrong}}{{Bernstein} \&
  {Armstrong}}{2014}]{ba14}
{Bernstein} G.~M.,  {Armstrong} R.,  2014, \mn@doi [\mnras]
  {10.1093/mnras/stt2326}, \href
  {http://adsabs.harvard.edu/abs/2014MNRAS.438.1880B} {438, 1880}

\bibitem[\protect\citeauthoryear{{Bernstein} \& {Jarvis}}{{Bernstein} \&
  {Jarvis}}{2002}]{bj02}
{Bernstein} G.~M.,  {Jarvis} M.,  2002, \mn@doi [\aj] {10.1086/338085}, \href
  {http://adsabs.harvard.edu/abs/2002AJ....123..583B} {123, 583}

\bibitem[\protect\citeauthoryear{{Bernstein}, {Armstrong}, {Krawiec}  \&
  {March}}{{Bernstein} et~al.}{2016}]{bfd2016}
{Bernstein} G.~M.,  {Armstrong} R.,  {Krawiec} C.,   {March} M.~C.,  2016,
  \mn@doi [\mnras] {10.1093/mnras/stw879}, \href
  {http://adsabs.harvard.edu/abs/2016MNRAS.459.4467B} {459, 4467}

\bibitem[\protect\citeauthoryear{{Bertin}}{{Bertin}}{2011}]{BertinPSFEx2011}
{Bertin} E.,  2011, in {Evans} I.~N.,  {Accomazzi} A.,  {Mink} D.~J.,   {Rots}
  A.~H.,  eds,  Astronomical Society of the Pacific Conference Series Vol. 442,
  Astronomical Data Analysis Software and Systems XX. p.~435

\bibitem[\protect\citeauthoryear{{Clowe}, {Brada{\v c}}, {Gonzalez},
  {Markevitch}, {Randall}, {Jones}  \& {Zaritsky}}{{Clowe}
  et~al.}{2006}]{CloweDMProof06}
{Clowe} D.,  {Brada{\v c}} M.,  {Gonzalez} A.~H.,  {Markevitch} M.,  {Randall}
  S.~W.,  {Jones} C.,   {Zaritsky} D.,  2006, \mn@doi [\apjl] {10.1086/508162},
  \href {http://adsabs.harvard.edu/abs/2006ApJ...648L.109C} {648, L109}

\bibitem[\protect\citeauthoryear{{Fenech Conti}, {Herbonnet}, {Hoekstra},
  {Merten}, {Miller}  \& {Viola}}{{Fenech Conti} et~al.}{2017}]{KidsShear2017}
{Fenech Conti} I.,  {Herbonnet} R.,  {Hoekstra} H.,  {Merten} J.,  {Miller} L.,
    {Viola} M.,  2017, \mn@doi [\mnras] {10.1093/mnras/stx200}, \href
  {http://adsabs.harvard.edu/abs/2017MNRAS.467.1627F} {467, 1627}

\bibitem[\protect\citeauthoryear{{Heymans} et~al.,}{{Heymans}
  et~al.}{2013}]{HeymansTomography2013}
{Heymans} C.,  et~al., 2013, \mn@doi [\mnras] {10.1093/mnras/stt601}, \href
  {http://adsabs.harvard.edu/abs/2013MNRAS.432.2433H} {432, 2433}

\bibitem[\protect\citeauthoryear{{Hirata}}{{Hirata}}{2016}]{HirataCorrNoise}
{Hirata} C.,  2016, in preparation

\bibitem[\protect\citeauthoryear{{Hirata} et~al.,}{{Hirata}
  et~al.}{2004}]{HirataAlign04}
{Hirata} C.~M.,  et~al., 2004, \mn@doi [\mnras]
  {10.1111/j.1365-2966.2004.08090.x}, \href
  {http://adsabs.harvard.edu/abs/2004MNRAS.353..529H} {353, 529}

\bibitem[\protect\citeauthoryear{{Hirata}, {Mandelbaum}, {Ishak}, {Seljak},
  {Nichol}, {Pimbblet}, {Ross}  \& {Wake}}{{Hirata}
  et~al.}{2007}]{HirataIntrinsicAlign07}
{Hirata} C.~M.,  {Mandelbaum} R.,  {Ishak} M.,  {Seljak} U.,  {Nichol} R.,
  {Pimbblet} K.~A.,  {Ross} N.~P.,   {Wake} D.,  2007, \mn@doi [\mnras]
  {10.1111/j.1365-2966.2007.12312.x}, \href
  {http://adsabs.harvard.edu/abs/2007MNRAS.381.1197H} {381, 1197}

\bibitem[\protect\citeauthoryear{{Hoekstra} \& {Jain}}{{Hoekstra} \&
  {Jain}}{2008}]{HoekstraJain2008}
{Hoekstra} H.,  {Jain} B.,  2008, \mn@doi [Annual Review of Nuclear and
  Particle Science] {10.1146/annurev.nucl.58.110707.171151}, \href
  {http://adsabs.harvard.edu/abs/2008ARNPS..58...99H} {58, 99}

\bibitem[\protect\citeauthoryear{{Hoekstra}, {Viola}  \&
  {Herbonnet}}{{Hoekstra} et~al.}{2017}]{Hoekstra2017}
{Hoekstra} H.,  {Viola} M.,   {Herbonnet} R.,  2017, \mn@doi [\mnras]
  {10.1093/mnras/stx724}, \href
  {http://adsabs.harvard.edu/abs/2017MNRAS.468.3295H} {468, 3295}

\bibitem[\protect\citeauthoryear{{Huff} \& {Mandelbaum}}{{Huff} \&
  {Mandelbaum}}{2017}]{HuffMcal}
{Huff} E.,  {Mandelbaum} R.,  2017, preprint, \href
  {http://adsabs.harvard.edu/abs/2017arXiv170202600H} {} (\mn@eprint {arXiv}
  {1702.02600})

\bibitem[\protect\citeauthoryear{{Huterer}, {Takada}, {Bernstein}  \&
  {Jain}}{{Huterer} et~al.}{2006}]{HutererSystematics06}
{Huterer} D.,  {Takada} M.,  {Bernstein} G.,   {Jain} B.,  2006, \mn@doi
  [\mnras] {10.1111/j.1365-2966.2005.09782.x}, \href
  {http://adsabs.harvard.edu/abs/2006MNRAS.366..101H} {366, 101}

\bibitem[\protect\citeauthoryear{{Ivezic} et~al.,}{{Ivezic}
  et~al.}{2008}]{IvezicLSST08}
{Ivezic} Z.,  et~al., 2008, preprint, \href
  {http://adsabs.harvard.edu/abs/2008arXiv0805.2366I} {} (\mn@eprint {arXiv}
  {0805.2366})

\bibitem[\protect\citeauthoryear{{Jarvis} et~al.}{{Jarvis}
  et~al.}{2016}]{Jarvis2016}
{Jarvis} M.,  et~al., 2016, \mn@doi [\mnras] {10.1093/mnras/stw990}, \href
  {http://adsabs.harvard.edu/abs/2016MNRAS.460.2245J} {460, 2245}

\bibitem[\protect\citeauthoryear{{Jee}, {Tyson}, {Hilbert}, {Schneider},
  {Schmidt}  \& {Wittman}}{{Jee} et~al.}{2016}]{Jee16}
{Jee} M.~J.,  {Tyson} J.~A.,  {Hilbert} S.,  {Schneider} M.~D.,  {Schmidt} S.,
   {Wittman} D.,  2016, \mn@doi [\apj] {10.3847/0004-637X/824/2/77}, \href
  {http://adsabs.harvard.edu/abs/2016ApJ...824...77J} {824, 77}

\bibitem[\protect\citeauthoryear{{Johnston} et~al.,}{{Johnston}
  et~al.}{2007}]{JohnstonLensing07}
{Johnston} D.~E.,  et~al., 2007, preprint, \href
  {http://adsabs.harvard.edu/abs/2007arXiv0709.1159J} {} (\mn@eprint {arXiv}
  {0709.1159})

\bibitem[\protect\citeauthoryear{{Kaiser}}{{Kaiser}}{2000}]{Kaiser2000}
{Kaiser} N.,  2000, \mn@doi [\apj] {10.1086/309041}, \href
  {http://adsabs.harvard.edu/abs/2000ApJ...537..555K} {537, 555}

\bibitem[\protect\citeauthoryear{{Kaiser}, {Squires}  \& {Broadhurst}}{{Kaiser}
  et~al.}{1995}]{ksb95}
{Kaiser} N.,  {Squires} G.,   {Broadhurst} T.,  1995, \apj, 449, 460+

\bibitem[\protect\citeauthoryear{{Kilbinger} et~al.,}{{Kilbinger}
  et~al.}{2013}]{CFHTCosmicShear2013}
{Kilbinger} M.,  et~al., 2013, \mn@doi [\mnras] {10.1093/mnras/stt041}, \href
  {http://adsabs.harvard.edu/abs/2013MNRAS.430.2200K} {430, 2200}

\bibitem[\protect\citeauthoryear{{Lackner} \& {Gunn}}{{Lackner} \&
  {Gunn}}{2012}]{LacknerGunn2012}
{Lackner} C.~N.,  {Gunn} J.~E.,  2012, \mn@doi [\mnras]
  {10.1111/j.1365-2966.2012.20450.x}, \href
  {http://adsabs.harvard.edu/abs/2012MNRAS.421.2277L} {421, 2277}

\bibitem[\protect\citeauthoryear{{Laureijs} et~al.,}{{Laureijs}
  et~al.}{2011}]{Euclid2011}
{Laureijs} R.,  et~al., 2011, preprint, \href
  {http://adsabs.harvard.edu/abs/2011arXiv1110.3193L} {} (\mn@eprint {arXiv}
  {1110.3193})

\bibitem[\protect\citeauthoryear{{Mandelbaum}, {Seljak}, {Kauffmann}, {Hirata}
  \& {Brinkmann}}{{Mandelbaum} et~al.}{2006}]{Mandelbaum06}
{Mandelbaum} R.,  {Seljak} U.,  {Kauffmann} G.,  {Hirata} C.~M.,   {Brinkmann}
  J.,  2006, \mnras, 368, 715

\bibitem[\protect\citeauthoryear{{Mandelbaum} et~al.,}{{Mandelbaum}
  et~al.}{2014}]{great3}
{Mandelbaum} R.,  et~al., 2014, \mn@doi [\apjs] {10.1088/0067-0049/212/1/5},
  \href {http://adsabs.harvard.edu/abs/2014ApJS..212....5M} {212, 5}

\bibitem[\protect\citeauthoryear{{Melchior} \& {Viola}}{{Melchior} \&
  {Viola}}{2012}]{Melchior12}
{Melchior} P.,  {Viola} M.,  2012, \mn@doi [\mnras]
  {10.1111/j.1365-2966.2012.21381.x}, \href
  {http://adsabs.harvard.edu/abs/2012MNRAS.424.2757M} {424, 2757}

\bibitem[\protect\citeauthoryear{{Melchior}, {Sutter}, {Sheldon}, {Krause}  \&
  {Wandelt}}{{Melchior} et~al.}{2014}]{MelchiorVoids2014}
{Melchior} P.,  {Sutter} P.~M.,  {Sheldon} E.~S.,  {Krause} E.,   {Wandelt}
  B.~D.,  2014, \mn@doi [\mnras] {10.1093/mnras/stu456}, \href
  {http://adsabs.harvard.edu/abs/2014MNRAS.440.2922M} {440, 2922}

\bibitem[\protect\citeauthoryear{{Miller} et~al.,}{{Miller}
  et~al.}{2013}]{Miller13}
{Miller} L.,  et~al., 2013, \mn@doi [\mnras] {10.1093/mnras/sts454}, \href
  {http://adsabs.harvard.edu/abs/2013MNRAS.429.2858M} {429, 2858}

\bibitem[\protect\citeauthoryear{{Moffat}}{{Moffat}}{1969}]{Moffat1969}
{Moffat} A.~F.~J.,  1969, \aap, \href
  {http://adsabs.harvard.edu/abs/1969A%26A.....3..455M} {3, 455}

\bibitem[\protect\citeauthoryear{{Noll}}{{Noll}}{1976}]{noll1976}
{Noll} R.~J.,  1976, Journal of the Optical Society of America (1917-1983),
  \href {http://adsabs.harvard.edu/abs/1976JOSA...66..207N} {66, 207}

\bibitem[\protect\citeauthoryear{{Okura} \& {Futamase}}{{Okura} \&
  {Futamase}}{2016}]{Okura2016}
{Okura} Y.,  {Futamase} T.,  2016, \mn@doi [\apj]
  {10.3847/0004-637X/827/2/138}, \href
  {http://adsabs.harvard.edu/abs/2016ApJ...827..138O} {827, 138}

\bibitem[\protect\citeauthoryear{{Refregier} \& {Amara}}{{Refregier} \&
  {Amara}}{2014}]{Refregier13}
{Refregier} A.,  {Amara} A.,  2014, \mn@doi [Physics of the Dark Universe]
  {10.1016/j.dark.2014.01.002}, \href
  {http://adsabs.harvard.edu/abs/2014PDU.....3....1R} {3, 1}

\bibitem[\protect\citeauthoryear{{Refregier}, {Kacprzak}, {Amara}, {Bridle}  \&
  {Rowe}}{{Refregier} et~al.}{2012}]{Refreg12}
{Refregier} A.,  {Kacprzak} T.,  {Amara} A.,  {Bridle} S.,   {Rowe} B.,  2012,
  \mn@doi [\mnras] {10.1111/j.1365-2966.2012.21483.x}, \href
  {http://adsabs.harvard.edu/abs/2012MNRAS.425.1951R} {425, 1951}

\bibitem[\protect\citeauthoryear{{Rowe} et~al.,}{{Rowe}
  et~al.}{2015}]{GALSIM2015}
{Rowe} B.~T.~P.,  et~al., 2015, \mn@doi [Astronomy and Computing]
  {10.1016/j.ascom.2015.02.002}, \href
  {http://adsabs.harvard.edu/abs/2015A%26C....10..121R} {10, 121}

\bibitem[\protect\citeauthoryear{{Schneider}, {Hogg}, {Marshall}, {Dawson},
  {Meyers}, {Bard}  \& {Lang}}{{Schneider}
  et~al.}{2015}]{SchneiderProbshear2015}
{Schneider} M.~D.,  {Hogg} D.~W.,  {Marshall} P.~J.,  {Dawson} W.~A.,  {Meyers}
  J.,  {Bard} D.~J.,   {Lang} D.,  2015, \mn@doi [\apj]
  {10.1088/0004-637X/807/1/87}, \href
  {http://adsabs.harvard.edu/abs/2015ApJ...807...87S} {807, 87}

\bibitem[\protect\citeauthoryear{{Scoville} et~al.,}{{Scoville}
  et~al.}{2007a}]{Scoville2007a}
{Scoville} N.,  et~al., 2007a, \mn@doi [\apjs] {10.1086/516580}, \href
  {http://adsabs.harvard.edu/abs/2007ApJS..172...38S} {172, 38}

\bibitem[\protect\citeauthoryear{{Scoville} et~al.,}{{Scoville}
  et~al.}{2007b}]{Scoville2007b}
{Scoville} N.,  et~al., 2007b, \mn@doi [\apjs] {10.1086/516751}, \href
  {http://adsabs.harvard.edu/abs/2007ApJS..172..150S} {172, 150}

\bibitem[\protect\citeauthoryear{{Seitz} \& {Schneider}}{{Seitz} \&
  {Schneider}}{1997}]{SeitzSchneider97}
{Seitz} C.,  {Schneider} P.,  1997, \aap, \href
  {http://adsabs.harvard.edu/abs/1997A%26A...318..687S} {318, 687}

\bibitem[\protect\citeauthoryear{{Sheldon} et~al.,}{{Sheldon}
  et~al.}{2009}]{SheldonLensing09}
{Sheldon} E.~S.,  et~al., 2009, \mn@doi [\apj] {10.1088/0004-637X/703/2/2217},
  \href {http://adsabs.harvard.edu/abs/2009ApJ...703.2217S} {703, 2217}

\bibitem[\protect\citeauthoryear{{Spergel} et~al.,}{{Spergel}
  et~al.}{2015}]{SpergelWFIRST2015}
{Spergel} D.,  et~al., 2015, preprint, \href
  {http://adsabs.harvard.edu/abs/2015arXiv150303757S} {} (\mn@eprint {arXiv}
  {1503.03757})

\bibitem[\protect\citeauthoryear{{Suchyta} et~al.,}{{Suchyta}
  et~al.}{2016}]{Balrog2016}
{Suchyta} E.,  et~al., 2016, \mn@doi [\mnras] {10.1093/mnras/stv2953}, \href
  {http://adsabs.harvard.edu/abs/2016MNRAS.457..786S} {457, 786}

\bibitem[\protect\citeauthoryear{{Takada}}{{Takada}}{2010}]{TakadaHSC2010}
{Takada} M.,  2010, in {Kawai} N.,  {Nagataki} S.,  eds,  American Institute of
  Physics Conference Series Vol. 1279, American Institute of Physics Conference
  Series. pp 120--127, \mn@doi{10.1063/1.3509247}

\bibitem[\protect\citeauthoryear{{The Dark Energy Survey Collaboration}}{{The
  Dark Energy Survey Collaboration}}{2005}]{DESWhitePaper}
{The Dark Energy Survey Collaboration} 2005, ArXiv Astrophysics e-prints, \href
  {http://adsabs.harvard.edu/abs/2005astro.ph.10346T} {}

\bibitem[\protect\citeauthoryear{{Troxel} \& {Ishak}}{{Troxel} \&
  {Ishak}}{2015}]{TroxelIAReview2015}
{Troxel} M.~A.,  {Ishak} M.,  2015, \mn@doi [\physrep]
  {10.1016/j.physrep.2014.11.001}, \href
  {http://adsabs.harvard.edu/abs/2015PhR...558....1T} {558, 1}

\bibitem[\protect\citeauthoryear{{Tyson}, {Valdes}, {Jarvis}  \&
  {Mills}}{{Tyson} et~al.}{1984}]{Tyson84}
{Tyson} J.~A.,  {Valdes} F.,  {Jarvis} J.~F.,   {Mills} A.~P.,  1984, \apjl,
  281, L59

\bibitem[\protect\citeauthoryear{{Zhang}}{{Zhang}}{2008}]{Zhang2008FourierQuadI}
{Zhang} J.,  2008, \mn@doi [\mnras] {10.1111/j.1365-2966.2007.12585.x}, \href
  {http://adsabs.harvard.edu/abs/2008MNRAS.383..113Z} {383, 113}

\bibitem[\protect\citeauthoryear{{Zhang}, {Luo}  \& {Foucaud}}{{Zhang}
  et~al.}{2015}]{Zhang2015}
{Zhang} J.,  {Luo} W.,   {Foucaud} S.,  2015, \mn@doi [\jcap]
  {10.1088/1475-7516/2015/01/024}, \href
  {http://adsabs.harvard.edu/abs/2015JCAP...01..024Z} {1, 024}

\bibitem[\protect\citeauthoryear{{Zhang}, {Zhang}  \& {Luo}}{{Zhang}
  et~al.}{2017}]{Zhang2017}
{Zhang} J.,  {Zhang} P.,   {Luo} W.,  2017, \mn@doi [\apj]
  {10.3847/1538-4357/834/1/8}, \href
  {http://adsabs.harvard.edu/abs/2017ApJ...834....8Z} {834, 8}

\bibitem[\protect\citeauthoryear{{Zuntz}, {Kacprzak}, {Voigt}, {Hirsch}, {Rowe}
   \& {Bridle}}{{Zuntz} et~al.}{2013}]{Zuntz13}
{Zuntz} J.,  {Kacprzak} T.,  {Voigt} L.,  {Hirsch} M.,  {Rowe} B.,   {Bridle}
  S.,  2013, \mn@doi [\mnras] {10.1093/mnras/stt1125}, \href
  {http://adsabs.harvard.edu/abs/2013MNRAS.434.1604Z} {434, 1604}

\makeatother
\end{thebibliography}

\end{document}